\documentclass{IEEEoj}
\usepackage{cite}
\usepackage{amsmath,amssymb,amsfonts}
\usepackage{algorithmic}
\usepackage{graphicx,color}
\usepackage{caption}
\usepackage{subcaption}
\usepackage{textcomp}
\usepackage{multicol}
\usepackage{float}
\usepackage{xcolor}
\usepackage{stfloats}  % Wichtig für figure* am Seitenboden
\usepackage[normalem]{ulem}
\usepackage{orcidlink}
\def\BibTeX{{\rm B\kern-.05em{\sc i\kern-.025em b}\kern-.08em
    T\kern-.1667em\lower.7ex\hbox{E}\kern-.125emX}}
\AtBeginDocument{\definecolor{ojcolor}{cmyk}{0.93,0.59,0.15,0.02}}
\def\OJlogo{\vspace{-14pt}\includegraphics[height=28pt]{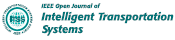}}
\begin{document}
\receiveddate{24 November, 2024}
\reviseddate{27 February, 2025}
\accepteddate{XX Month, XXXX}
\publisheddate{XX Month, XXXX}
\currentdate{XX Month, XXXX}
\doiinfo{OJITS.2022.1234567}

\title{Evaluation of Remote Driver Performance in Urban Environment Operational Design Domains}

\author{OLE HANS\textsuperscript{1,2}\orcidlink{0009-0009-0239-9469}, BENEDIKT WALTER\textsuperscript{2}\orcidlink{0009-0002-4213-4421} and JÜRGEN ADAMY\textsuperscript{1}\orcidlink{0000-0001-5612-4932}}
\affil{Department of Control Methods and Intelligent Systems, Technical University of Darmstadt, Darmstadt, 64289 Germany}
\affil{Department of Operational Safety, Vay Technology GmbH, Berlin, 
12099 Germany}
\corresp{CORRESPONDING AUTHOR: OLE HANS (e-mail: olehans@outlook.de).}
\authornote{The work of Ole Hans was supported by the Operational Safety Department of Vay Technology GmbH.}
\markboth{Evaluation of Remote Driver Performance in Urban Environment Operational Design Domains}{Ole Hans \textit{et al.}}

\begin{abstract}
Remote driving has emerged as a solution for enabling human intervention in scenarios where Automated Driving Systems (ADS) face challenges, particularly in urban Operational Design Domains (ODDs). This study evaluates the performance of Remote Drivers (RDs) of passenger cars in a representative urban ODD in Las Vegas, focusing on the influence of cumulative driving experience and targeted training approaches. Using performance metrics such as efficiency, braking, acceleration, and steering, the study shows that driving experience can lead to noticeable improvements of RDs and demonstrates how experience up to 600 km correlates with improved vehicle control. In addition, driving efficiency exhibited a positive trend with increasing kilometers, particularly during the first 300 km of experience, which reaches a plateau from 400 km within a range of 0.35 to \mbox{0.42 km/min} in the defined ODD. The research further compares ODD-specific training methods, where the detailed ODD training approaches attains notable advantages over other training approaches. The findings underscore the importance of tailored ODD training in enhancing RD performance, safety, and scalability for Remote Driving System (RDS) in real-world applications, while identifying opportunities for optimizing training protocols to address both routine and extreme scenarios. The study provides a robust foundation for advancing RDS deployment within urban environments, contributing to the development of scalable and safety-critical remote operation standards.
\end{abstract}


\maketitle

\section{INTRODUCTION}
\IEEEPARstart{A}{utomated} Driving Systems (ADS) have the potential to revolutionize transportation by increasing efficiency, safety, and passenger comfort, as well as reducing mobility's economic and environmental costs \cite{litman2020autonomous, schoitsch2016autonomous}. However, ADS face significant challenges, particularly in urban environments where interaction with other traffic participants such as pedestrians or other vehicles is inevitable. One prominent issue involves stranded vehicles situations in which ADS encounter complex scenarios beyond their programmed capabilities, often requiring human intervention to navigate. This issue is further complicated by the necessity for ADSs to interpret and respond to unpredictable human behavior, posing a substantial hurdle to full autonomy and public acceptance of such systems \cite{graf2020improving, mutzenich2021updating, cooke2006human}.

The concept of remote operation, specifically remote driving, has emerged as a viable solution to these challenges, enabling human Remote Drivers (RD) to take control of ADS in situations where ADS fall short. By leveraging RD, they can support ADS to resolve complex scenarios more effectively, improving the overall reliability, availability and safety of automated operations in complex Operational Design Domains (ODDs). Remote driving can thereby act as a bridge toward safer integration of ADS in urban environments, complementing the technological capabilities of ADS with the adaptability and intuition of human operators \cite{hans2024backedautonomy}.

The implementation of remote driving, however, comes with its own set of stringent requirements. The BSI Flex 1887 \cite{BSIFlex1887} standard, for example, outlines essential training protocols for RD, emphasizing the need for familiarity with various vehicle states and specific ODD conditions, as well as realistic, training environments. Despite these defined standards, there remains a lack of comprehensive research to validate the effectiveness of these training requirements in real-world conditions. Existing studies on RD performance have often relied on simulators \cite{neumeier2019teleoperation} or controlled environments \cite{den2022design, tener2022driving} with untrained participants, which fail to capture the full complexity of real-world urban driving. Simulator studies are valuable in understanding general human responses at different speeds, but cannot reproduce the absolute validity of real driving situations \cite{hussain2019speed}.

This gap between industrial applications and scientific validation is underscored by the proliferation of remote driving implementations in industry by companies such as Vay \cite{vay_teledriving}, Fernride \cite{fernride2024safety}, or DriveU \cite{DriveU2024}. These companies demonstrate RD's practical potential, however, scientific research into human performance within real-world ODDs remains sparse. The Federal Highway Research Institute (BASt) has also recognized this research deficit, calling for further study into effective training methods for remote operators in its publication on research needs in teleoperation \cite{TeleoperationReport}.

Moreover, the integration of remote driving in urban mobility ecosystems also poses challenges in public perception. According to a recent McKinsey study \cite{Heineke2024} on urban mobility, concerns regarding the coexistence of remote-controlled and automated vehicles with human-driven vehicles on public roads persist. As consumer readiness for such innovations remains uncertain, the need for transparent research and effective training standards for RDs in real-world ODDs becomes all the more pressing. 

The novelty of the study is to address the outlined research gap by evaluating RD performance in passenger cars in urban ODDs under real-world conditions, contributing to a more robust and scientifically grounded foundation for RD in intelligent transportation systems. This work addresses the gap in the scientific investigation of RDS and offers the opportunity to verify assumptions on human performance in real ODDs based on simulation studies. In the first part of the study it is to be investigated to what extent remote driving experience influence the performance and driving efficiency of RDs and whether the requirements from \mbox{BSI Flex 1887 \cite{BSIFlex1887}} can be implemented and validated in practice. In the second part the study analyses different training approaches for ODDs to compare different training approaches for RDs within specific ODDs, focusing on identifying which training method best enhances controllability, efficiency, and scalability to achieve optimal performance of the RDS.

Therefore, Section \ref{relatedwork} provides literature research relevant to the study, and Section \ref{AnalysisDataPreparation} defines the key metrics used in this study. Section \ref{SYSTEMMODEL} describes the RDS in its parts and the respective ODD as the basis for this study. The results of investigation of driving experience on RD performance are described in Section \ref{Investigation}, followed by the analysis about different RD training approaches for ODDs in Section \ref{EVALUATION OF ODD-SPECIFIC}. The limitations of the results are discussed in Section \ref{Limitation}. Finally, the conclusion and potential further work are presented in Section \ref{conclusion}. 

\section{RELATED WORK}
\label{relatedwork}
This section presents an overview of research related to remote driving, the ODD, and the assessment of driving performance.

\subsection{Remote Driving}
\label{remotedriving}
Remote driving in the automotive industry is distinguished from related concepts like remote assistance and remote monitoring, forming a nuanced taxonomy of Remote Human Input Systems that varies in complexity \cite{UNECE, bogdoll2022taxonomy, amador2022survey}. Remote driving involves the direct control of a vehicle by a RD, from a remote location, enabling the navigation of a vehicle through complex environments without requiring physical presence. Building on foundational work by \mbox{Bogdoll et al. \cite{bogdoll2022taxonomy}} and Amador et al. \cite{amador2022survey}, this field is classified by levels of complexity, which aligns with the SAE taxonomy for ADS \cite{SAEI}. An overview of which remote operation approaches are best suited to safely and efficiently return ADS to automated operation can be found in the systematic analysis by \mbox{Brecht et al. \cite{brecht2024evaluation}}.

\begin{figure}[!h]
\centering
  \includegraphics[width=0.35\textwidth]{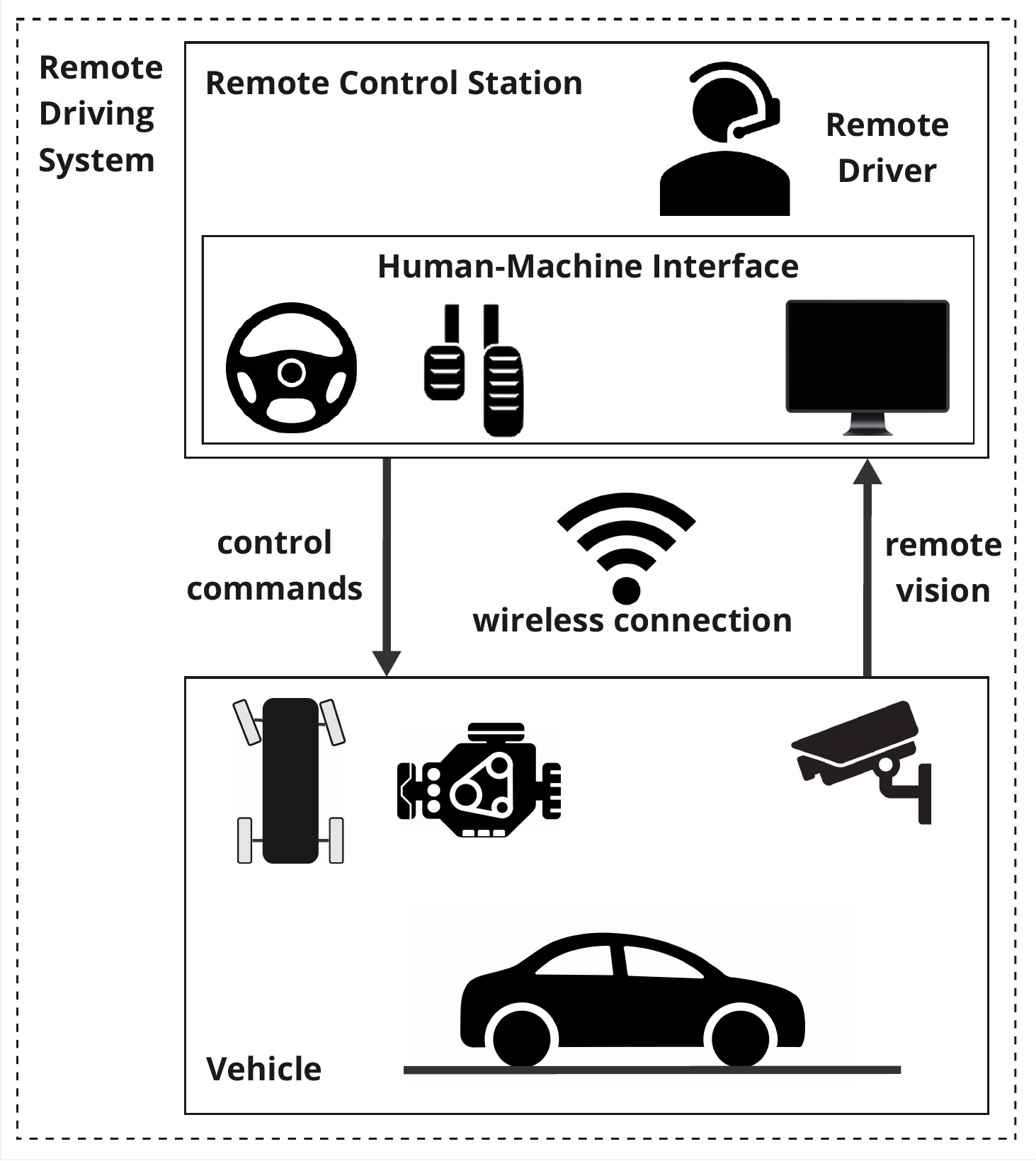}
    \caption{Simplified visualization of a Remote Driving System (RDS).}
    \label{RDSOverview}
\vspace{-3mm}
\end{figure}

The core of remote driving lies in a stable wireless connection between the Remote Control Station (RCS) and the vehicle, enabling real-time monitoring and control (see Fig. \ref{RDSOverview}). This connection allows the RD to access parts of the ODD remotely, using data transmitted via cameras and further sensors, albeit with latency that impacts real-time the RD's situational awareness \cite{kettwich2021teleoperation, neumeier2019teleoperation}. The approach to control can be either direct, where the RD continuously monitors and reacts to the environment, or indirect, where intervention occurs only if the ADS encounters operational limits \cite{kettwich2021teleoperation}.

Challenges in RDS operation are both technical and human-centered. Technical challenges such as latency, video quality, and visibility impairments create performance constraints for the system itself as well as for the RD, while human factors, such as the lack of haptic feedback, reduce the RD's situational awareness \cite{chen2007human, hortal2019rehabilitation, hans2024human}. The RD relies solely on visual feedback, which poses unique limitations compared to conventional driving. Latency and reduced video quality can further impede the RD’s capacity to maintain consistent situational awareness while operating the vehicle \cite{tener2022driving, neumeier2019teleoperation}.

In light of these limitations, a well-defined ODD must reflect not only system capabilities but also the human operator’s performance threshold to ensure effective RD functionality \cite{hans2023operational}. Addressing these challenges demands both functional adaptations and operational measures, emphasizing the need for both technological enhancements and targeted human performance training to mitigate the inherent limitations of remote driving \cite{Schwindt-Drews2024, hans2024human}. 

Current remote driving research literature lacks of work based on public street data, highlighting that most existing studies are conducted in simulators \cite{neumeier2019teleoperation} or controlled environments \cite{den2022design, tener2022driving} with untrained participants, which fail to capture the full complexity of real-world urban driving. This study contributes by identifying key factors that should be considered when transitioning remote driving operations to real-world settings.

\subsection{Operational Design Domain}
The ODD defines the system's specific operational scope within the broader Operational Domain (OD), encompassing the operating conditions applicable to the system \cite{rohne2022implementing}. ODDs play a crucial role in system safety assessment and validation, ensuring that the system operates within predefined safe and predictable conditions \cite{sun2021acclimatizing}. When system requirements derived from the ODD can no longer be met, either due to system limitations or failures, a fallback for the Dynamic Driving Task (DDT) must be activated, initiating an active degradation function or transitioning to a Minimal Risk \mbox{Condition (MRC)} to mitigate risks during the \mbox{transition \cite{SAEI, no2021157}}. An MRC represents a lower risk condition and is triggered by a Minimum Risk \mbox{Maneuver (MRM)}, a core safety function to maintain control within the ODD limits.

The ODD's parameters typically include factors such as road geometry, weather and time of day, along with considerations of connectivity and human-machine \mbox{interactions \cite{czarnecki2018towards}}. Given the wide array of potential conditions, the ODD definition often relies on simplifications, with established classifications offering systematic criteria for describing ODDs \cite{PAS, koopman2019many, wachenfeld2016safety, AVSC, FDIS}. Various sources use the abbreviation ODD to provide requirements for an ODD \cite{UL4600}, guidelines for testing and validation of autonomous vehicles \cite{IAMTS}, and to address ODD in conjunction with other topics such as Safety of the Intended Functionality, also known as SOTIF \cite{SOTIF}. 

ODD definitions also serve to delineate system boundaries, which constrain the scope of safety cases and testing efforts. Testing within a defined ODD enables realistic system validation, fault identification, and performance optimization. Scenario-based validation strategies, developed for specific ODDs, allow testing of critical scenarios related to vehicle functions. This approach reduces efforts by focusing on scenario coverage rather than extensive distance-based testing \cite{king2020taxonomy}. Using scenario-based methods, ODDs can be tested within controlled environments or simulated with high levels of detail, capturing a range of real-world interactions.

By providing a structured approach to validation, ODD definitions and scenario-based testing frameworks ensure that systems can safely operate and respond to variable conditions, maintaining alignment with defined operational boundaries and enhancing reliability in both highway and urban environments.
For remotely driven vehicles in urban environment, a comprehensive ODD qualification process was defined by Hans et al. \cite{hans2023operational}.

\begin{figure*}
\centerline{\includegraphics[width=\textwidth]{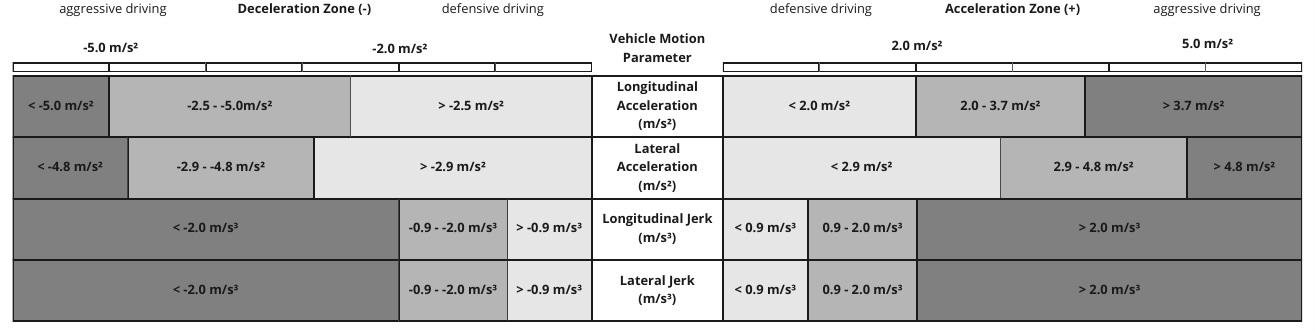}}
\caption{Longitudinal and lateral vehicle motion parameters under consideration.
\label{motionpara}}
\vspace{-3mm}
\end{figure*}

\subsection{Driving Performance Measuring}
\label{Drivingmetrics}
A driver's driving style is understood as the way the driver operates the vehicle controls in the context of the driving scene and external conditions such as time in the day, day of the week, or weather \cite{martinez2017driving}. The driving style can be categorized by a spectrum of behavioral and vehicle dynamics that reflect driver preferences and tendencies, such as speed selection, acceleration patterns, and risk-taking \mbox{behavior \cite{elander1993behavioral}}. Research has shown that driving styles commonly fall into three main clusters: comfort-oriented, moderate, and dynamic/aggressive. For instance, \mbox{Bellem \cite{bellem2018comfort}} categorizes drivers into comfort-focused, normal, and aggressive driving style. Abendroth and Bruder \cite{abendroth2009leistungsfahigkeit} define driving styles based on metrics like speed, longitudinal acceleration, and following distance, describing drivers as slow and comfort-conscious, average with high safety awareness, and fast and sporty. This aligns with Schulz and Frömings \cite{schulz2008analyse} definition, who define categories such as calm-adaptive, active-dynamic, sporty-ambitious, affective-imbalanced, uncertain-clumsy, and aggressive-reckless driving styles. Summarizing these classifications, most research divides driving styles into defensive, normal, and aggressive categories, each with associated characteristics. A literature review by Tselentis and Apadimitriou \cite{tselentis2023driver} confirms aggressive driving is a common driving pattern classification. Several other authors used these categories and note that other definitions can fit within this \mbox{framework \cite{buyukyildiz2017identification, karjanto2017simulating, martinez2017driving}}. Distinct patterns in longitudinal and lateral acceleration, as shown by \mbox{Pion et al. \cite{pion2012fingerprint}}, further support the correlation between driving style and vehicle dynamics, where drivers with aggressive styles show more extreme acceleration values. Therefore, identifying driving style involves a synthesis of both behavioral intentions and dynamic vehicle indicators, where driving style classifications support various applications. 

To identify the driving style, the relevant variables that need to be monitored to enable robust classification need to be determined. However, the literature lacks consensus on a standard set of parameters, partly due to the diverse applications of driving style measurements, such as driver correction, fuel economy improvement, and safety enhancement \cite{ericsson2000variability, taubman2004multidimensional}. Simpler approaches focus on the relationship between aggressive driving styles and high acceleration, speed, and fuel consumption \cite{corti2013quantitative, manzoni2010driving, guardiola2014modelling}. These models typically include metrics such as speed, acceleration, deceleration, and incorporate braking patterns \cite{xu2015establishing, lenaers2009real}. 

Martinez et al. \cite{martinez2017driving} provides a holistic overview about possible metrics which can be used for this. The data, on which the choice of these driving metrics is based, comes from various sources that study typical human driving styles under real-world conditions. These studies provide insights into the lateral and longitudinal dynamic movements of vehicles, which are influenced by driving style, and enable an in-depth analysis of driving styles \cite{hugemann2003longitudinal, bogdanovic2013research, el2007evaluation, le2015autonomous}. The measurement of lateral acceleration was also identified as an important indicator of driving style, as it provides information about the cornering behavior and stability of the vehicle \cite{schulz2008analyse}. Longitudinal dynamic parameters on the other hand, reflect the driver's control over the vehicle in terms of speed and deceleration \cite{karjanto2017simulating}. Additionally, advanced indicators, like jerk (rate of change in acceleration), throttle pressure, and braking frequency, are valuable for capturing sudden and aggressive driving behaviors \cite{murphey2009driver, miyajima2007driver}. The combination of longitudinal and lateral dynamics, as suggested by Doshi and Trivedi \cite{doshi2010examining}, allows for a more comprehensive capture of driving style.

The potential for increased steering aggression and abrupt jerks in remote driving scenarios, which can negatively impact passenger comfort, has been highlighted in recent studies. Papaioannou et al. \cite{papaioannou2023motion} found that remote driving scenarios often increase motion sickness among passengers. These findings underscore the need for further research on how remote driving impacts specific aspects of driving behavior, such as lateral and longitudinal precision, which are crucial for driving style measurement in remote and automated driving contexts. While previous studies have explored various aspects of driving performance, there remains a significant gap in empirical assessments of RD driving performance in real-world traffic environments. Therefore, this study uses real-world remote driving data to evaluate the driving performance of the RD.

\section{Analysis of Driving Metrics and Data Preparation}
\label{AnalysisDataPreparation}
Driving metrics provide an essential foundation for assessing the performance, safety, and efficiency of RDs in RDS. This section introduces the key metrics used in this study, their thresholds for harsh and extreme events, and the methodology employed for data generation and processing. The metrics analyzed include acceleration, deceleration, lateral acceleration, jerk, and efficiency, as they are fundamental to understanding driving styles and vehicle dynamics.

\begin{table}[h!]
\centering
\begin{tabular}{|l|l|l|}
\hline
\textbf{Event Type} & \textbf{Threshold} & \textbf{Reference} \\ \hline
Braking events       & $< -2.5 \, \text{m/s}^2$       & \cite{schulz2008analyse, buyukyildiz2017identification, schwab2019methode, lee2024study} \\ \hline
Acceleration events  & $> 2.0 \, \text{m/s}^2$       & \cite{karjanto2017simulating, mayser2004fahrerassistenzsysteme, lee2024study, svensson2015tuning, bosetti2014human} \\ \hline
Right steering events & $> 2.9 \, \text{m/s}^2$      & \cite{hugemann2003longitudinal, dorr2014online, buyukyildiz2017identification, schwab2019methode} \\ \hline
Left steering events  & $< -2.9 \, \text{m/s}^2$     & \cite{hugemann2003longitudinal, dorr2014online, buyukyildiz2017identification, schwab2019methode} \\ \hline
Deceleration Jerk events & $< -0.9 \, \text{m/s}^3$ & \cite{murphey2009driver, svensson2015tuning} \\ \hline
Acceleration Jerk events & $> 0.9 \, \text{m/s}^3$  & \cite{kilinc2012determination, bae2019toward} \\ \hline
Right lateral jerk events & $> 0.9 \, \text{m/s}^3$ & \cite{bae2019toward} \\ \hline
Left lateral jerk events  & $< -0.9 \, \text{m/s}^3$ & \cite{bae2019toward} \\ \hline
\end{tabular}
\caption{Vehicle motion parameters, thresholds, and corresponding references for identifying relevant events.}
\label{motionparajournal}
\vspace{-3mm}
\end{table}

\subsection{Driving Metrics and Event Classification}
In this study, key driving metrics, including acceleration, braking, and lateral acceleration, were analyzed due to their critical relevance for understanding human driving styles and their impact on vehicle dynamics. Additionally, the variable for longitudinal and lateral jerk, representing the rate of change of acceleration, was included. The focus of the analysis was limited to harsh and extreme driving events, representing critical behaviors with extreme parameter values. Tab. \ref{motionparajournal} summarizes the vehicle motion parameters, their thresholds based on Fig. \ref{motionpara}, and corresponding references.

To evaluate the temporal and spatial efficiency of RDs, Remote Driving Efficiency (RDE) was assessed using kilometers driven per minute as shown in \eqref{eq1}. This metric offers insights into the RD's performance by quantifying the relationship between remotely driven distance and time.

\begin{equation}
\label{eq1}
\text{RDE} = \frac{\text{remotely driven distance (km)}}{\text{remotely driven time (min)}}
\end{equation}

Efficiency trends across time intervals were analyzed to determine whether improvements correlate with increasing driving experience. Such findings could provide important information for the optimization of training programs and the adaptation of operational strategies.

\subsection{Data Generation and Event Aggregation}
During remote driving sessions, the RDS in this study continuously collects the relevant data every second, capturing vehicle dynamics and status, including position, velocity, and acceleration parameters. The telemetry system transmits this information via proprietary vehicle physics messages, enabling real-time insights into the vehicle dynamics.

Specific driving events as defined in Tab. \ref{motionparajournal} are identified and aggregated if they persist across multiple seconds. Events occurring within a five-second window are combined, and the resulting event duration is calculated. This aggregation approach enhances the precision of driving style metrics and facilitates the analysis of extended driving scenarios.

\subsection{Data Preparation and Filtering}
The dataset underwent rigorous preparation and filtering to ensure the reliability and validity of the analysis. The following criteria were applied:

\begin{itemize}
    \item \textbf{Analysis period:} The analysis covered the period from August 1, 2023, to October 30, 2024, ensuring sufficient data for trend identification.
    \item \textbf{Remote driving data only:} Data from conventional driving sessions was excluded to maintain focus on remote driving performance.
    \item \textbf{Session length exclusion:} Sessions shorter than \mbox{0.1 m} were excluded to avoid incomplete or irrelevant data points due to performed vehicle start-up checks. 
    \item \textbf{System Under Test exclusion:} Data recorded under specific test conditions of the Vay system was removed to avoid biases.
    \item \textbf{ODD compliance:} Only data from within the defined ODD, as specified in Section IV.\ref{RDS ODD}, was considered.
    \item \textbf{Public roads only:} Data from parking lots or restricted testing areas were excluded.
    \item \textbf{Driver roles:} Data was filtered to include only designated RD (as defined in Section IV.\ref{RDTraining}). Additionally, driving sessions under specific conditions, such as those involving a safety driver were considered to ensure the relevance and consistency of the dataset. Sessions where no individual was present in the vehicle were also included, as they reflect scenarios of full remote operation, ensuring the analysis captured all relevant use cases.
\end{itemize}

This meticulous data preparation ensures that the analysis focuses solely on real-world driving scenarios under consistent and defined conditions. By excluding potential confounding factors, the study aims to derive robust insights into RD performance and driving efficiency.

\section{REMOTE DRIVING SYSTEM MODEL}
\label{SYSTEMMODEL}
In this study, the RDS developed and operated in a Las Vegas ODD of Vay Technology was utilized. The system consists of three primary components: the vehicle, which is equipped with the Vay hardware and software, the RCS, which enables remote control, and the human-in-the-loop, the RD.

The vehicle, a Kia eNiro, was retrofitted to integrate the remote driving technology, including additional cameras and in-house developed safety controller that monitor and regulate critical safety parameters in real-time. Furthermore, the vehicle’s connectivity was enhanced by specific antennas and modems, integrated with the proprietary Vay connectivity software stack, ensuring a stable and fast communication link between the vehicle and the RCS.

\begin{figure}[ht]
\centerline{\includegraphics[width=2.5in]{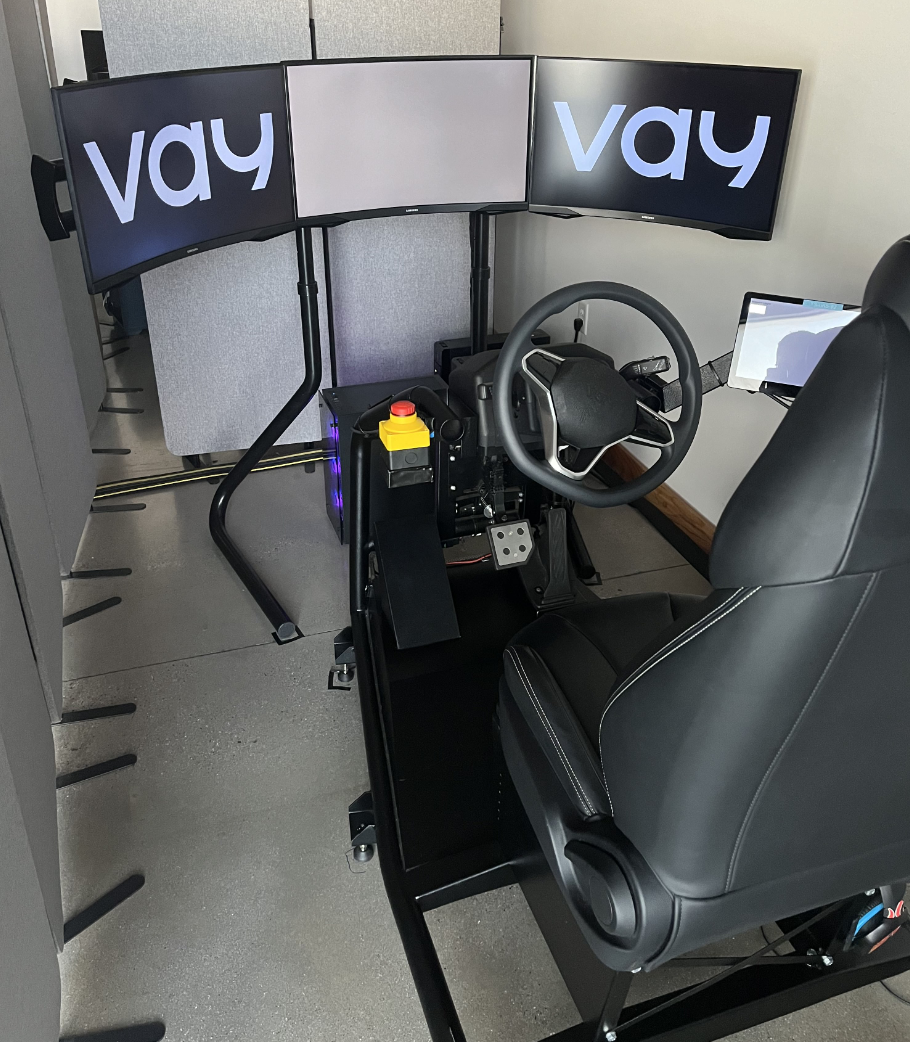}}
\caption{Vay Remote Control Station (RCS) within an operations center in Las Vegas, Nevada.
\label{VayRCS}}
\vspace{-3mm}
\end{figure}

The RCS, shown in Fig. \ref{VayRCS}, serves as the Human-Machine-Interface for the RD, who operates the vehicle remotely. It features three screens that display visual feedback from the vehicle's cameras and further sensors. Additionally, speakers offer auditory feedback and communication to the inside of the vehicle. Visual perception is transmitted through multiple camera sensors, while road traffic sounds are delivered via external microphones to the RD's headphone. The vehicle is controlled via a automotive-grade physical steering wheel, along with automotive-grade controls such as column switches, throttle, and brake pedals. Special controllers in the RCS process incoming data and enable interaction with the Vay system installed in the vehicle.

\subsection{Remote Driving System Operational Design Domain}
\label{RDS ODD}
The choice of the described ODD ensures that the RDS guarantees the required connectivity and controllability within a defined environment \cite{hans2024backedautonomy}. These framework conditions are intended to create the basis for the safe control and monitoring of the RDS using the human-in-the-loop approach. The ODD of the RDS is specified with regard to the environment, traffic structure, speed limits, weather conditions and time of day. The RDS used for this study operates exclusively in an urban environment, specifically Las Vegas, Nevada. The urban environment offers a variety of traffic and infrastructure conditions that the system has to cope with. However, the ODD excludes specific parameters such as:

\begin{itemize}
    \item \textbf{Speed limitation:} Streets in the defined ODD are limited to a maximum of \mbox{35 mph}. Therefore, highways and interstates are explicitly not included, as usually have a higher speed limit than \mbox{35 mph}.
    \item \textbf{Weather conditions:} Specific conditions such as snow, ice and rain are excluded from the ODD in this study which are achieved almost continuously by the Las Vegas city location.
    \item \textbf{Time of day:} The use of RDS is limited to driving during daylight hours in this study.
    \item \textbf{Sufficient and stable connectivity:} As \mbox{Hans et al. \cite{hans2023operational}} point out in their methodology for the ODD definition, connectivity is an essential prerequisite for the operation of the RDS within the described ODD and must be sufficiently guaranteed for the reliable exchange of information of the driving environment. Accordingly, areas such as tunnels, are checked for sufficient connectivity and often excluded due to the connectivity problems that frequently occur there. A stable communication link is a key ODD requirement for the safe operation of the RDS, which has been identified as one of the main limitations in previous studies \cite{hans2024backedautonomy}.
    \item \textbf{Human in the loop:} In contrast to ADS applications, the RDS does not require the system itself to take over the driving task completely, but instead relies on a human-in-the-loop approach. The task of the RD therefore requires specific qualifications that focus primarily on the safe control of the vehicle in the defined context of the ODD.
\end{itemize}

The precise definition of the ODD, including specifications on the environment, speed, weather, connectivity and RD training, should ensure that the RDS can be operated safely and reliably in the defined area of application.

\begin{figure*}[ht] % Use figure* to span both columns
    \centering
    \begin{subfigure}{0.325\textwidth}
        \includegraphics[width=\linewidth]{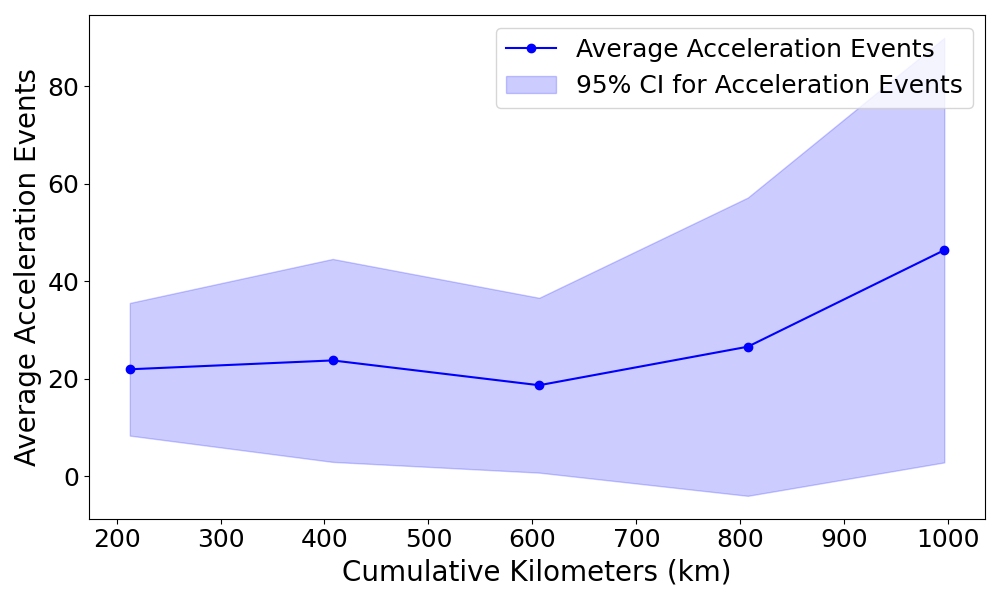}
        \caption{Average Acceleration Events}
        \label{Average Acceleration Events}
    \end{subfigure}
    \hfill
    \begin{subfigure}{0.325\textwidth}
        \includegraphics[width=\linewidth]{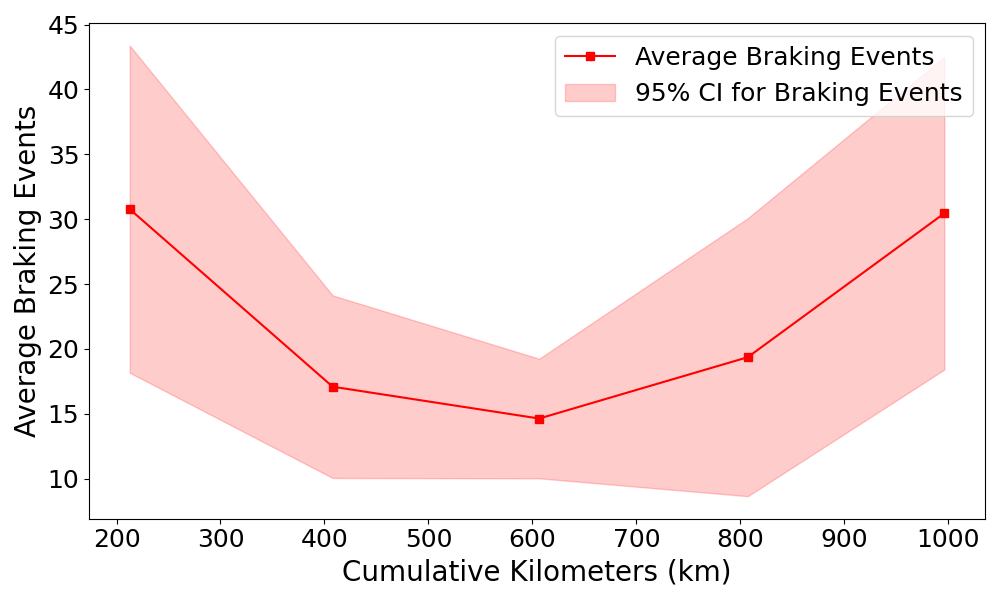}
        \caption{Average Braking Events}
        \label{Average Braking Events}
    \end{subfigure}
    \hfill
    \begin{subfigure}{0.325\textwidth}
        \includegraphics[width=\linewidth]{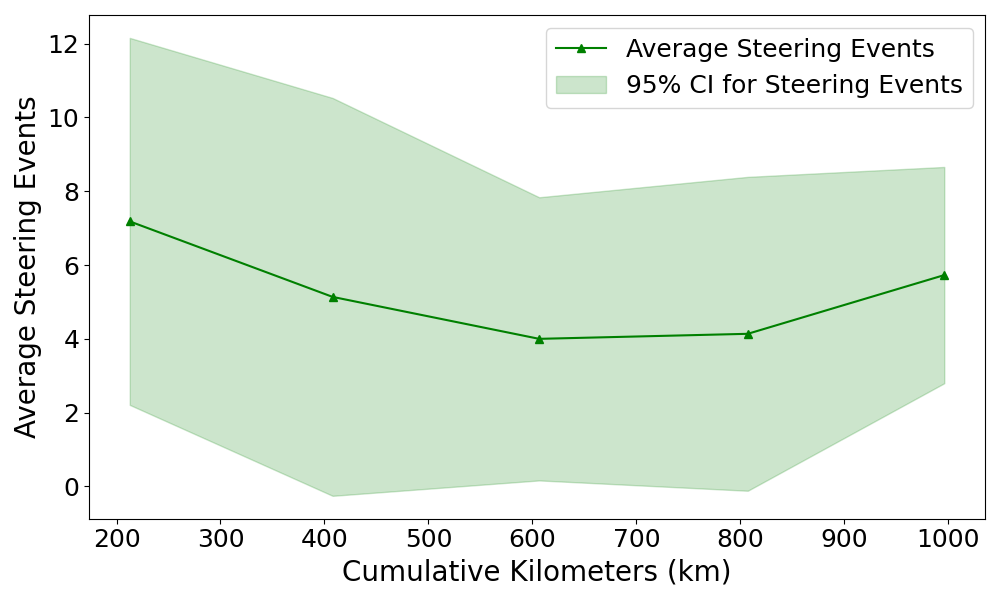}
        \caption{Average Steering Events}
        \label{Average Steering Events}
    \end{subfigure}
    \begin{subfigure}{0.325\textwidth}
        \includegraphics[width=\linewidth]{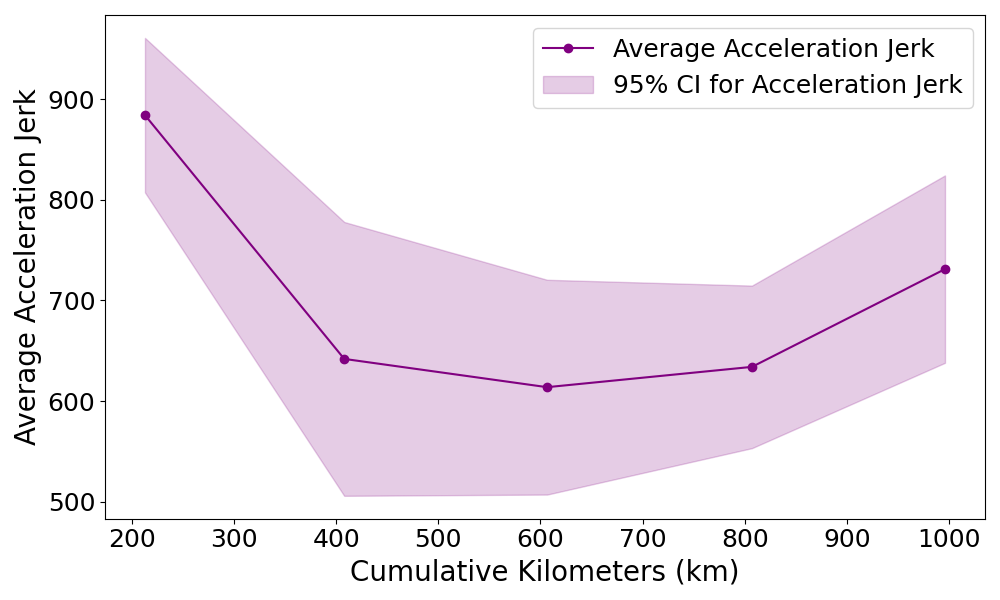}
        \caption{Average Acceleration Jerk Events}
        \label{Average Acceleration Jerk Events}
    \end{subfigure}
    \hfill
    \begin{subfigure}{0.325\textwidth}
        \includegraphics[width=\linewidth]{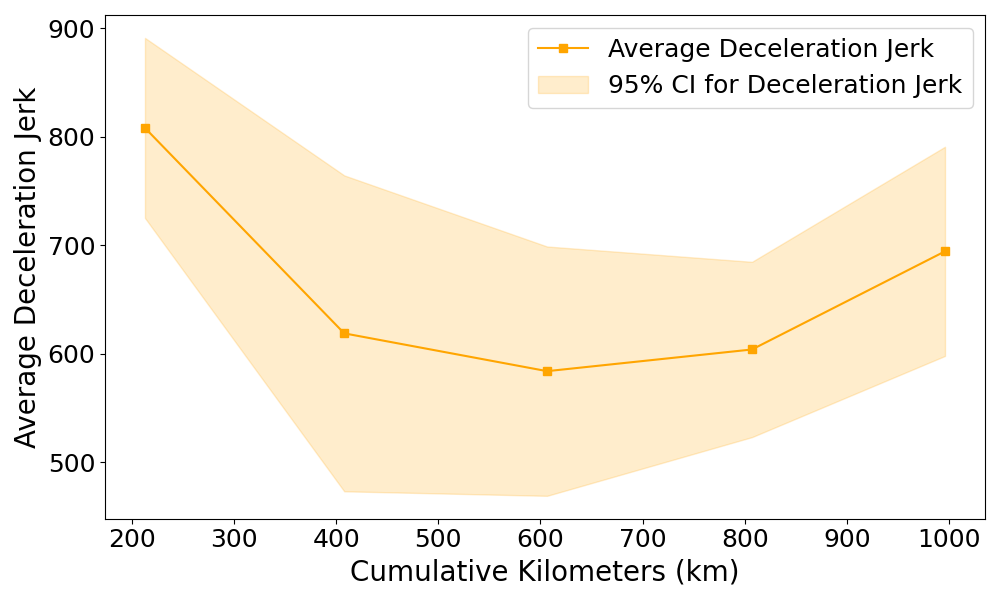}
        \caption{Average Deceleration Jerk Events}
        \label{Average Deleration Jerk Events}
    \end{subfigure}
    \hfill
    \begin{subfigure}{0.325\textwidth}
        \includegraphics[width=\linewidth]{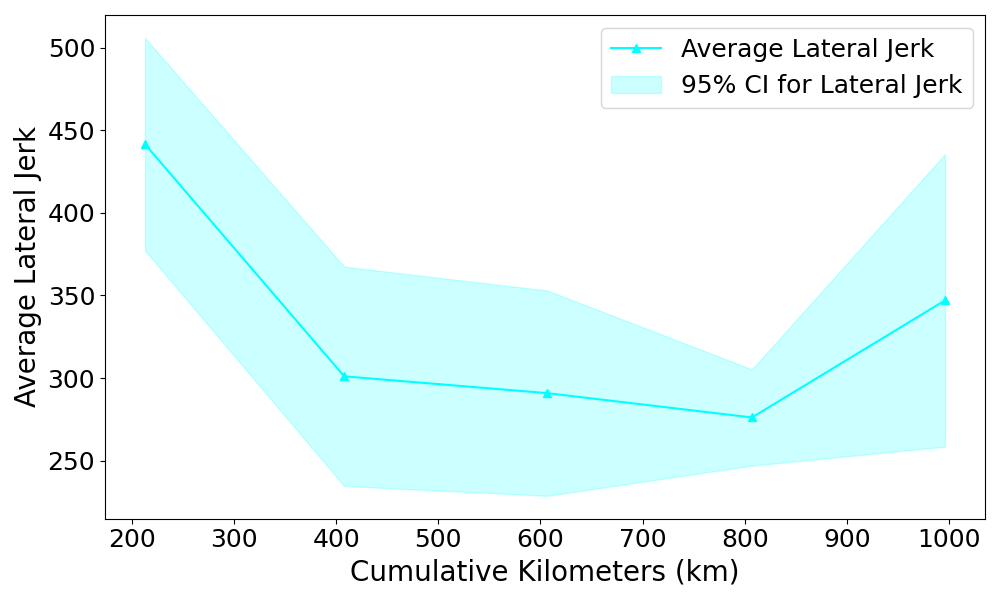}
        \caption{Average Lateral Jerk Events}
        \label{Average Lateral Jerk Events}
    \end{subfigure}
    \caption{Overview of average metrics by event type with 95{\%} confidence interval.}
    \label{Overview Average Metrics}
\end{figure*}

\subsection{Remote Driver Training}
\label{RDTraining}
To ensure the safe and effective use of an RDS in the defined ODD, specific training of RDs is necessary. Even though the RDs considered in this evaluation already have a drivers license and at least three years of driving experience, this is not sufficient to meet the unique requirements of remotely controlling a vehicle due to the inherent limitations of the RDS as described in Section II.\ref{remotedriving}. Scientific research emphasizes that the remote operation of the system and the response to vehicle-specific control requirements requires additional skills from experienced drivers \cite{hans2024human, Schwindt-Drews2024}. 

A structured training program that combines theoretical and practical content ensures that RDs are able to safely and proactively manage the RDS and prepare for situation-specific challenges in the ODD. The aim is to provide RDs with comprehensive and specific knowledge for the safe operation of the RDS. The training program of the RDs considered in this study has been described in detail by Hans \cite{Hans2023Academy} and ensures that the RDs are optimally prepared both technically and as drivers. The entire training program takes place with a Safety Driver as a fall-back level in the remotely driven vehicle.

\begin{itemize}
    \item \textbf{RDS-related RD training:} In the first phase of the training, the focus is on the operation and control of the RDS. The RDs are introduced to the specific functions and operating characteristics of the system in order to develop a deep understanding of its technical capabilities and limitations. This includes in particular the control, the adaption to latency and the adaptation to possible connection difficulties. The aim is for the RDs to gain a confident feel for the remote control of the vehicle and be able to make optimal use of the technology. The RDS-related RD training includes several theory classes with thematically related practical driving lessons. The first driving lessons take place on private training ground until the RD can demonstrate a certain level of driving competence and ability to control the RDS. This includes special driving maneuver training and getting used to safety concept such as the MRM concept, which can potentially occur. Only once this has been achieved, the RD can gain driving experience on public roads to control the system in complex real-world driving situations, with an additional human fallback level in the vehicle to intervene if necessary.
    \item \textbf{ODD-related RD training:} The second phase focuses on the understanding and practical application of the RDS within the defined ODD. An essential part of the ODD training is spatial orientation and understanding of the local environment. RDs learn to take into account the specifics of the urban environment, e.g. road sections with high traffic volumes, areas with potentially weak network connectivity or complex intersections. As stated by Hardin \cite{Hardin2024well}, RDs have noted that local knowledge of streets is essential to maintaining focus on the driving task, allowing them to anticipate hazardous areas that may involve higher road user density, inconsistent network connections, or complex intersections. This local awareness aids drivers in preparing for specific challenges within urban ODDs, enhancing their ability to manage safety-critical situations and ensuring smoother integration with other road users.
\end{itemize}

With the additional baseline assessment of the RD conducted prior to the training to determine the initial skill level and the standardized training duration and progression to ensure that all participants undergo comparable training scenarios, objective performance metrics, such as those presented in this study, can be used to evaluate when an RD is ready to remotely drive without an additional human fallback level, such as a safety driver.

\section{INVESTIGATION OF DRIVING EXPERIENCE ON REMOTE DRIVER PERFORMANCE}
\label{Investigation}
The primary objective is to assess how prior driving experience influences the driving performance of RDs within a specified ODD. This analysis focused on quantifying the relationship between experience and driving performance metrics with the RDS utilizing data collected from a controlled set of RDs ($n = 14$) operating in a representative ODD designed according to the ODD qualification process outlined by Hans et al. \cite{hans2023operational}. Of the 14 RD, aged between 23 and 31 years (\mbox{$M = 27.79$}, \mbox{$SD = 2.49$}), 12 identified themselves as male and 2 as female.

All RDs were selected based on their completion of standardized training, which they undertook on a closed course to simulate real-world scenarios under safe, controlled conditions. This prerequisite ensured a uniform level of initial competence across RDs, allowing the study to isolate the effects of experience on performance rather than variations in basic driving skills. To evaluate driving performance, the work focused on specific control and efficiency metrics defined in Section II.\ref{Drivingmetrics}.

Within a total remotely driven distance of 24,221.5 km, each RD was observed over a series of driving sessions in the defined ODD environment, with continuous data collection on the selected performance metrics. By the performance metrics, the study aimed to discern patterns linking driving experience with the ability to maintain stability, safety, and responsiveness in urban traffic situations.

\subsection{Driving Events per RD Driving Experience}
Fig. \ref{Overview Average Metrics} illustrates the average metrics for the defined event types across cumulative driving experiences, measured in kilometers, with 95\% confidence intervals. The trends observed across these metrics provide valuable insights into the learning and adaptation of RDs over time.

The results indicate that increasing driving experience correlates with a reduction in abrupt steering, braking, and acceleration actions, suggesting improved vehicle control and the development of driving habits, particularly in the early stages (up to \mbox{600 km}). This trend is visible across various metrics in Fig. \ref{Overview Average Metrics}, including lateral jerk, deceleration jerk, and acceleration jerk, all of which decrease substantially with experience before stabilizing or slightly increasing around the 800 to 1000 km range, likely due better driving experience, routine and development of driving habits. 

The observed decrease in Spearman-coefficient correlations between cumulative driving experience and different event types, illustrates a learning curve where increasing experience leads to improved driving stability and lower event frequency. For example, at 200 km, the correlations for braking events per date and longitudinal acceleration events per date are relatively high (0.37 and 0.41, respectively), indicating frequent and less controlled maneuvers in early driving phases. However, these correlations decrease at \mbox{500 km} and beyond. For example, braking events drop to 0.01 up to \mbox{500 km}  and longitudinal acceleration events to 0.08, indicating more consistent and anticipatory driving behavior.

Overall, it is important to note that the confidence intervals are relatively large, likely due to the small sample size $n$. This amplifies the impact of outliers, making the observed effects less stable. Despite this limitation, the trends are still evident, demonstrating the robustness of the observed patterns even within a limited data set.

\subsection{Efficiency per RD Driving Experience}
RDE is a useful metric for evaluating the performance of RDs, particularly as they gain experience over time. The analysis uses \eqref{eq1}, which defines efficiency as the ratio of remotely driven distance to remotely driven time. This metric provides a standardized measure of driving performance, allowing comparisons across varying levels of experience. By assessing efficiency at incremental stages of cumulative experience, the study aims to identify trends and potential inflection points in RD performance, offering insights the role of experience in achieving optimal driving outcomes.

\begin{figure*}[ht] % Use figure* to span both columns
    \centering
    \begin{subfigure}{0.325\textwidth}
        \includegraphics[width=\linewidth]{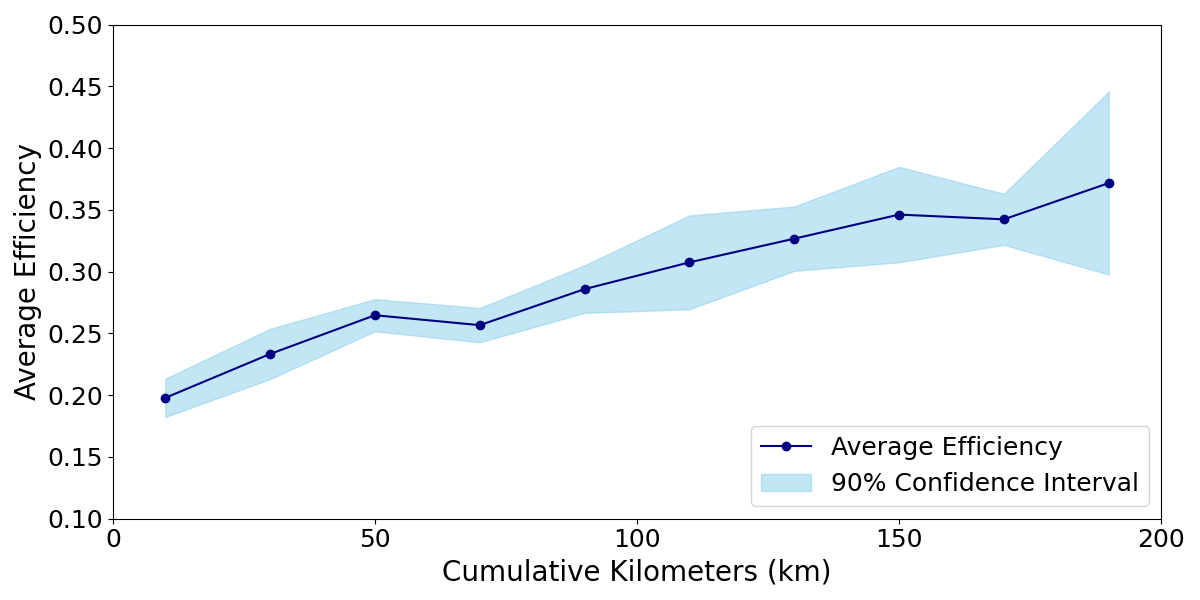}
        \caption{Efficiency until 200km Remote Driving Experience}
    \label{Eff until 200km}
    \end{subfigure}
    \hfill
    \begin{subfigure}{0.325\textwidth}
        \includegraphics[width=\linewidth]{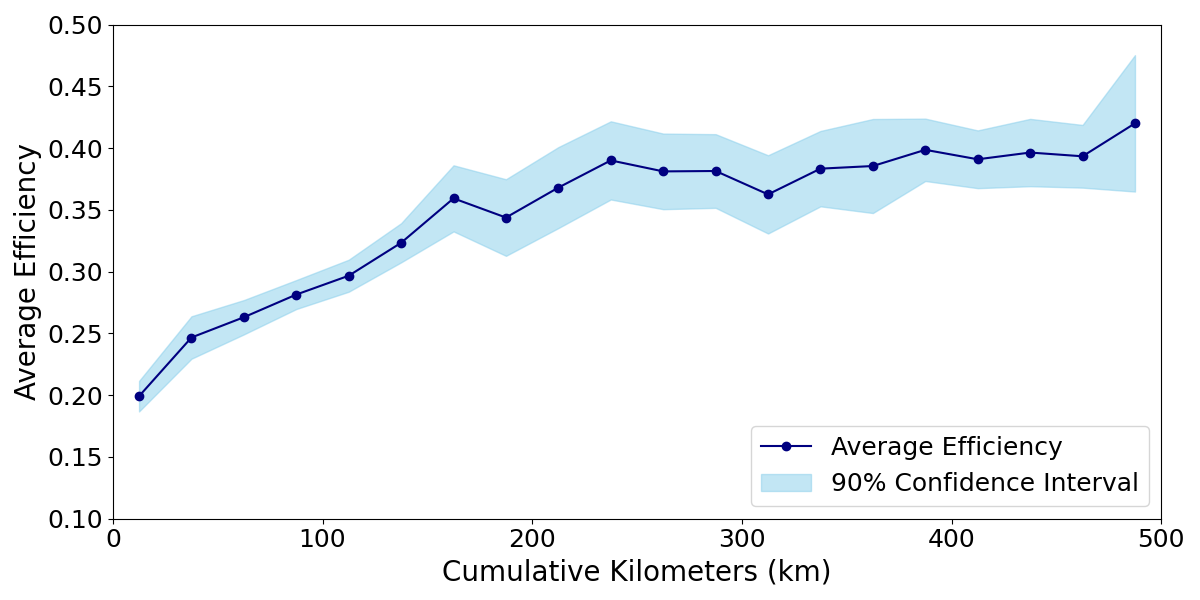}
        \caption{Efficiency until 500km Remote Driving Experience}
    \label{Eff until 500km}
    \end{subfigure}
    \hfill
    \begin{subfigure}{0.325\textwidth}
        \includegraphics[width=\linewidth]{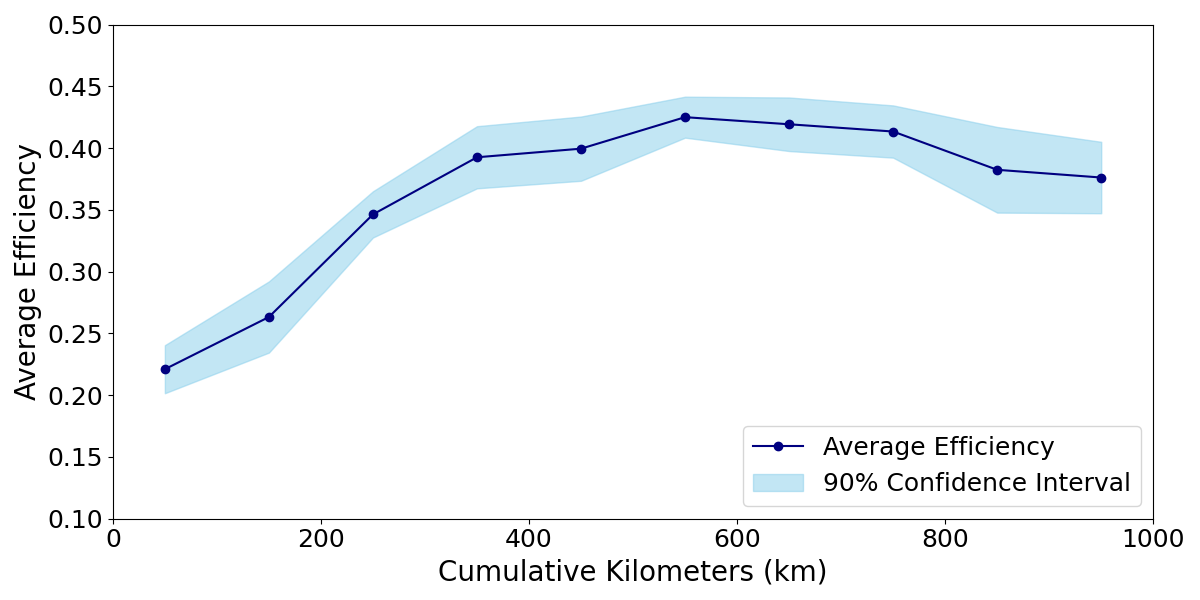}
        \caption{Efficiency until 1000km Remote Driving Experience}
    \label{Eff until 1000km}
    \end{subfigure}
    \caption{Efficiency per remote driving experience (200km, 500km, and 1000km) with 90\% confidence interval.}
    \label{Efficency}
    \vspace{-3mm}
\end{figure*}

\subsubsection{Efficiency Performance of RD until 200 km}
The efficiency of most RDs, visualized in Fig. \ref{Efficency}, improves with increasing driving experience. Both the efficiency of the RDs and their average efficiency generally improve as mileage increases. This suggests that experience in distance traveled has a positive influence on efficiency. This could indicate that the RDs adapt to the conditions of the ODD or become more consistent in their driving behavior, which contribute to the increased efficiency. At this early stage of driving experience, there is a moderately positive correlation between km/day and efficiency (0.57) and between min/day and efficiency (0.42), according to the used Spearman-coefficient. This indicates that drivers who drive more are more efficient. At this early stage of experience, regular practice plays an important role in increasing efficiency, as RDs improve through repeated rides. Both correlations are significant ($p < 0.01$), suggesting that these relationships are statistically significant and unlikely to be due to chance.

\subsubsection{Efficiency Performance of RD until 500 km}
While the efficiency increases significantly within or throughout the first \mbox{300 km}, the additional kilometers seem to have less influence on the efficiency. After about 400 km, the average efficiency of the RD settles at a certain plateau between 0.35 and 0.45 with an individual RD efficiency maximum of 0.5. This could indicate that additional kilometers only result in limited increases in efficiency. This suggests that after a certain amount of experience, RDs reach an optimal performance level in terms of efficiency for the defined ODD. RDs usually improve up to a certain point; after reaching this point, efficiency gains are limited. This points to the need to provide specific training to optimize performance beyond this point.

Until a total experience of 500 km, the correlation between km/day and efficiency remains positive (0.41), but is slightly lower than before. The correlation between min/day and efficiency also decreases to 0.24. This could indicate that with increasing driving experience, efficiency can be increased less strongly by daily mileage and driving time, as drivers become increasingly familiar with the system and the conditions. Both correlations remain statistically significant at $p < 0.01$, which supports the robustness of the relationship.

\subsubsection{Efficiency Performance of RD until 1000 km}
Up to 1000 km of total experience, the correlation between km/day and efficiency remains unchanged at 0.41, and the correlation between min/day and efficiency increases slightly to 0.26. Efficiency still appears to be boosted by regular driving at this stage, but to a lesser extent, as drivers have probably already reached a certain level of competence. Both correlations are significant at $p < 0.01$, indicating a reliable relationship between these variables, even if the strength of the correlation is relatively moderate. The efficiency drop at 900 km is difficult to interpret due to the large number of potential influencing factors as such outliers are more weighty due to the small sample size $n$.

\subsubsection{Individual RD Efficiency Performance}
There is a considerable variation both between RDs and within individual RD performance across driving sessions. This could be due to different driving conditions or other situational factors affecting each session. There are differences in efficiency between different RDs, which could indicate individual driving styles and individual adaptations to the operating conditions of an RDS. While some RDs show consistently high efficiencies, others show greater fluctuations, possibly indicating different responses to driving conditions. The high variability in efficiency between RDs suggests that personalized training approaches are required to address the individual strengths and weaknesses of drivers.

\begin{figure}[!h]
\centerline{\includegraphics[width=3.5in]{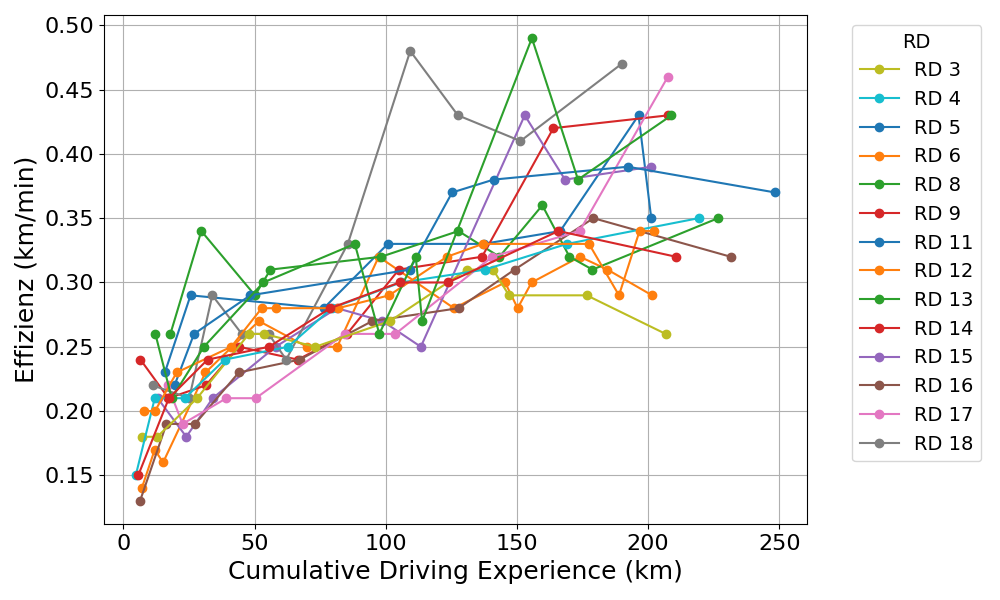}}
\caption{Individual efficiency performance of Remote Drivers (RD) until an experience level of 200 km.}
\label{RD Efficiency}
\vspace{-3mm}
\end{figure}

Furthermore, Fig. \ref{RD Efficiency} illustrates that some RDs, such as \mbox{RD 8} and \mbox{RD 14}, excel in efficiency, while other RDs, e.g. \mbox{RD 3}, are in the lower range of efficiency. These differences may reflect different training levels, driving techniques or individual responses to challenges while driving. Overall, the results show that the efficiency of the RD increases with experience in distance traveled, but not all RDs benefit equally from this experience.

\section{EVALUATION OF ODD-SPECIFIC TRAINING APPROACHES}
\label{EVALUATION OF ODD-SPECIFIC}
This part of the study will delve into how the ODD-specific training approach affects driving performance and to what extent specific ODD training approaches can improve driving performance and safety for RDs in addition to the general RD training evaluated in Section \ref{Investigation}. This Section compares different training approaches for RDs within ODDs, defined according to the ODD qualification process from \mbox{Hans et al. \cite{hans2023operational}}, focusing on identifying which training approach best enhances safety, efficiency, and scalability to achieve optimal remote driving performance. A key part will involve evaluating potential behavioral abnormalities as defined in Section II.\ref{Drivingmetrics} such as braking patterns or speed 
variations across different RD groups to uncover practical differences between the ODD training approaches. The driving metrics for longitudinal and lateral jerk are not considered. Additionally, the study will assess the cost-benefit ratio of each approach, with an emphasis on training scalability, safety and efficiency.

The ODD-specific training to be evaluated are clustered into the following three approaches: 

\begin{itemize}
\item \textbf{Approach 1: No specific ODD-based training} is provided to the RDs. Instead of focusing on the specific conditions or scenarios within a defined ODD, RDs are trained with a generalized skill set, as defined in Section IV.\ref{RDTraining} for remote driving without considering the constraints or challenges unique to particular ODDs. The assumption in this approach is that RDs possess the necessary skills to adapt to any driving environment or scenario without situation-specific ODD training.
\item \textbf{Approach 2: Detailed road section based ODD training approach} focuses on the thorough analysis and training of all possible conditions and parameters within a specific ODD including the geofence (see \mbox{Fig. \ref{fig2}}). The goal is for RDs to gain a deep understanding of each variable that may influence driving behavior. This includes environmental factors (e.g., weather conditions, time of day, road conditions) and especially traffic-related scenarios (e.g., worst-case and edge-case scenarios).
\item \textbf{Approach 3: Scenario-based ODD training approach} focuses, in contrast to the detailed approach, on training of the RDs by using representative driving scenarios that reflect typical challenges within an ODD. Rather than isolating all possible variables, this approach prepares drivers for real-world situations that are likely to encounter.
\end{itemize}

\subsection{Detailed vs. scenario-based ODD Training}
\label{detailedvsscenario}
The effectiveness of scenario-based ODD training versus detailed, road-specific ODD training was analyzed to assess their impact on driving safety and performance. This comparison provides critical insights for establishing scalable, safety-critical training methods for remote driving applications. Two distinct training approaches were evaluated:
\begin{itemize}
    \item \textbf{Detailed, road-specific ODD training:} RDs were trained over 5 months, from August 1 to \mbox{December 30,} 2023, covering a total of 1134.85 km. Of the 6 RD within this training group, aged between 26 and 45 years (\mbox{$M = 33.33$}, \mbox{$SD = 6.55$}), 6 identified themselves as male and 2 as female.
    \item \textbf{Scenario-based ODD training:} RDs trained over a longer period of 7 months, from January 1 to \mbox{July 30,} 2024, with a slightly higher total distance of \mbox{1163.43 km}. This group consisted of 8 RDs, aged between 24 and 33 years (\mbox{$M = 29.17$}, \mbox{$SD = 3.312$}), 6 identified themselves as male and 2 as female.
\end{itemize}

\begin{figure}[h!]
\centerline{\includegraphics[width=2.2in]{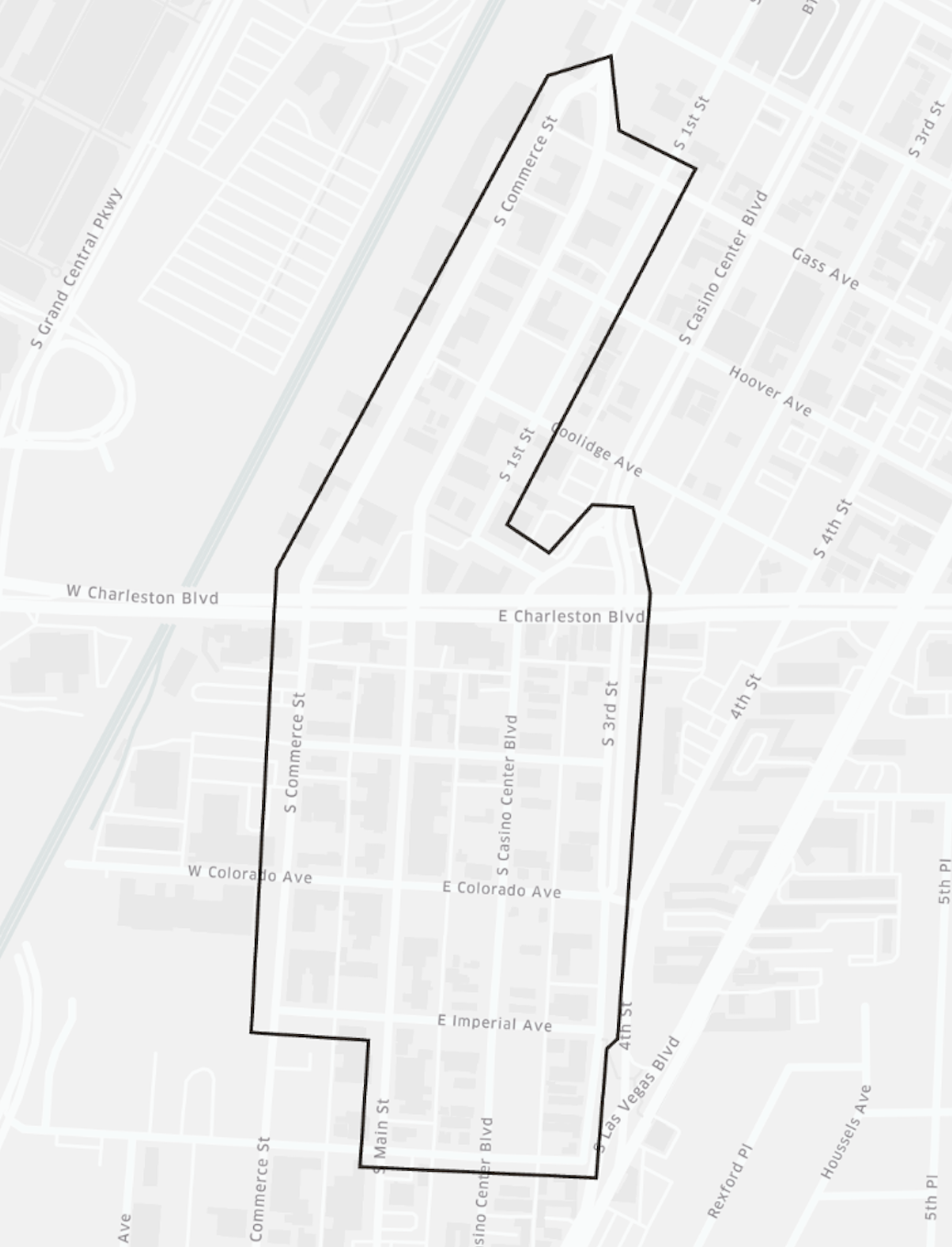}}
\caption{Geographical part of approach 2 and 3 Operational Design Domain (ODD) in Las Vegas, Nevada.\label{fig2}}
\vspace{-3mm}
\end{figure}

Despite differences in the training periods, both groups had comparable total distances in the geographical part of the ODD (see Fig. \ref{fig2}), ensuring consistency in exposure to driving conditions.

\subsubsection{Longitudinal Event Results}
The results of the longitudinal event performance analysis for RDs trained under approaches 2 and 3 reveal significant differences in braking behavior, stemming from the distinct training methodologies employed in each approach. Approach 2 fosters a more stable and anticipatory braking strategy, attributed to its intensive training across diverse scenarios, including edge and worst-case situations. In contrast, approach 3 encourages a reactive driving style, characterized by higher braking frequency and greater variability. These differences are reflected in the mean braking decelerations and the variability of braking behavior, both of which are supported by statistical analyses.

\begin{figure}
\centerline{\includegraphics[width=3.5in]{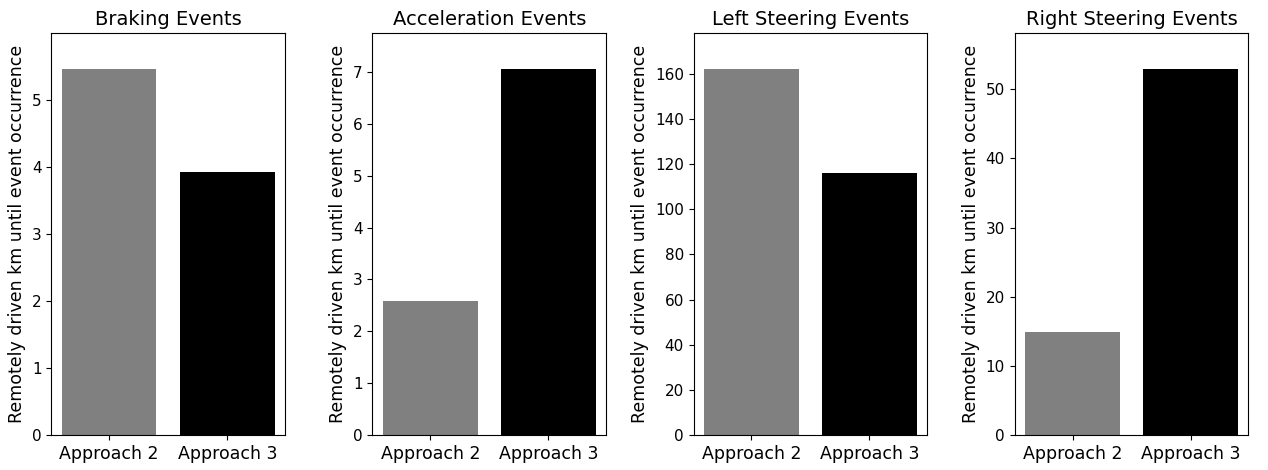}}
\caption{Remotely driven kilometers per event group until event occurrence for approach 2 and approach 3.\label{approach2kmperevent123}}
\vspace{-3mm}
\end{figure}

Drivers trained under approach 2 braked less frequently on average compared to approach 3 as shown in Fig.\ref{approach2kmperevent123}. This reduced frequency suggests that approach 2 drivers developed a more anticipatory and consistent braking style. Additionally, the mean braking deceleration of approach 2 RDs was -2.89 m/s$^2$, less severe than the more negative mean of -3.08 m/s$^2$ observed in approach 3, further highlighting the reactive nature of the latter group.

The standard deviation (SD) of braking decelerations also underscores these differences, with approach 2 showing a lower variability ($SD = 0.64$) compared to approach 3 ($SD = 0.96$). This suggests that the scenario-based training in approach 3 leads to more frequent and variable responses to traffic situations, including occasional extreme braking reactions. 

Statistical tests confirm these observations, with a Mann-Whitney U-test ($p = 0.001$) demonstrating significant differences in the distribution of braking events and Levene’s test ($p = 0.026$) indicating significant differences in variance between the two groups. 

\begin{figure}[!h]
\centerline{\includegraphics[width=3.5in]{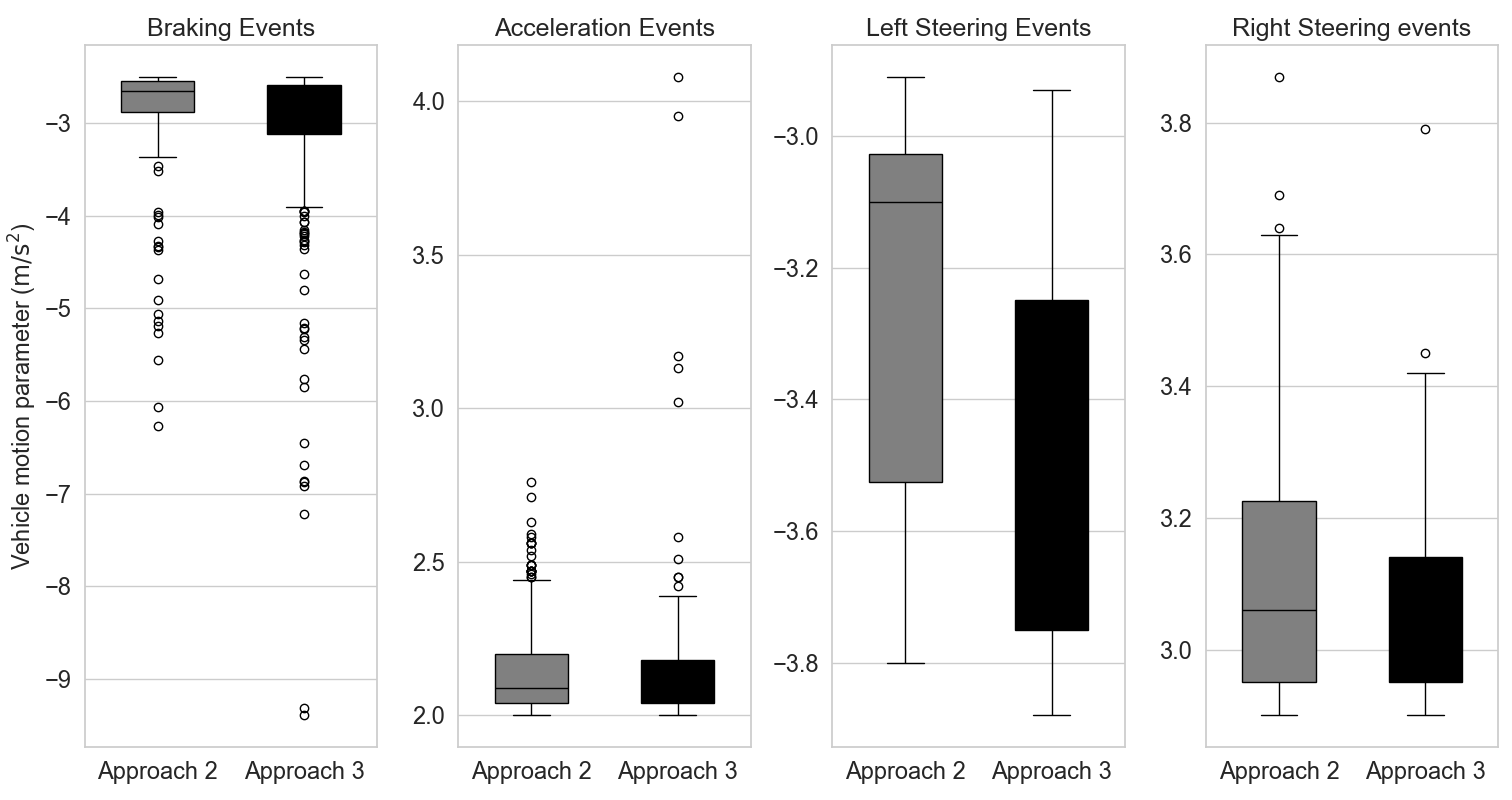}}
\caption{Driving performance boxplots per event type for approach 2 and approach 3.\label{approach2boxplotperevent123}}
\end{figure}

Furthermore, approach 3 exhibited more frequent and pronounced outliers in braking intensity as visualized in \mbox{Fig. \ref{approach2boxplotperevent123}}, suggesting occasional strong braking reactions prompted by unexpected situations.

While the significant differences in braking performance are evident, they may partly be explained by differences in RD driving experience between the two groups. A Mann-Whitney U-test analyzing driving experience ($<$ 500 km vs. $>$ 500 km) resulted in a p-value of $p = 0.0426$, indicating a statistically significant difference in experience distribution. Similarly, Levene’s test yielded a significant p-value ($p = 0.0016$), confirming that the variances between the two groups are not equal. This finding suggests that differences in braking behavior may be influenced, at least in part, by the varying levels of driving experience.

The analysis of acceleration events reveals nuanced differences in driving behavior between RDs trained under approaches 2 and 3 as well as some notable similarities. RDs in approach 3 accelerated more frequently as shown in \mbox{Fig. \ref{approach2kmperevent123}}. These findings suggest that drivers in approach 3 maintain a more consistent and less aggressive acceleration behavior. This observation aligns with the findings from Section \ref{Investigation} and is further supported by \mbox{Fig. \ref{Average Acceleration Events}}, taking into account the imbalance in driver experience illustrated in \mbox{Tab. \ref{Rdisbalnce}}.

\begin{table}[!h]
\centering
\begin{tabular}{p{45pt}p{65pt}p{75pt}}
\hline
RD experience & Detailed ODD-specific training & Scenario-based ODD- specific training \\ \hline \hline
$<$ 500 km & 256.23 km & 1059.85 km \\ \hline 
$>$ 500 km & 878.62 km & 103.57 km \\ \hline
\end{tabular}
\caption{Overview of Remote Driver (RD) experience for approach 2 (detailed ODD-specific training) and approach 3 (scenario-based ODD-specific training).}
\label{Rdisbalnce}
\vspace{-3mm}
\end{table}

Despite differences in acceleration frequency, the mean acceleration values were similar between the approaches, with approach 2 averaging 2.14 m/s$^2$ and approach 3 slightly higher at 2.16 m/s$^2$. However, the variability in acceleration was greater in approach 3 ($SD = 0.27$) compared to approach 2 ($SD = 0.14$), indicating a more dynamic response to varying traffic scenarios among approach 3 accelerations drivers. This variability could reflect the scenario-based training in approach 3, which emphasizes adaptability to diverse and unpredictable traffic situations.

The median values and distributions of acceleration forces were comparable across both approaches, demonstrating uniform patterns in acceleration. Statistical tests, including the Mann-Whitney U-test ($p = 0.764$) and Levene's test ($p = 0.183$), confirmed no significant differences in the distribution and variability of acceleration events between the two approaches. This suggests that, while approach 3 drivers may experience greater variability, the overall acceleration performance remains consistent with that of approach 2 drivers.

Approach 3, however, exhibits some stronger positive outliers, indicating occasional intense acceleration maneuvers. These outliers may be attributed to specific traffic scenarios requiring rapid responses, consistent with the reactive nature of the scenario-based training in approach 3. These findings collectively highlight that, while both approaches exhibit comparable acceleration behavior on average, the increased frequency and variability observed in approach 3 may reflect the influence of their distinct training methodology and driving scenarios.

\subsubsection{Lateral Event Results}
The lateral performance analysis of RDs trained under approaches 2 and 3 reveals differences in steering behavior, with notable variations in event frequency and consistency. Despite these differences, statistical tests largely indicate comparability in the distribution and variability of steering events between the two groups.

Approach 3 exhibited a higher average event rate for right-hand steering movements, with drivers averaging \mbox{50 km} per event, compared to 15 km per event for approach 2 as shown in \mbox{Fig. \ref{approach2kmperevent123}}. This suggests that approach 3 drivers engage in more frequent steering corrections, likely due to variations in driving patterns and environmental conditions encountered during training or operations. However, the mean steering accelerations for right-hand movements were similar between the two approaches, with \mbox{approach 2} recording \mbox{2.14 m/s$^2$} and approach 3 showing \mbox{2.16 m/s$^2$}. Variability in these movements was also comparable for \mbox{approach 2} ($SD = 0.21$) and approach 3 ($SD = 0.22$).

Statistical tests confirmed no significant differences between the two groups. The Mann-Whitney \mbox{U-test} yielded a p-value of $p = 0.689$, while Levene’s test reported $p = 0.942$, indicating that both the distribution and variability of right steering events are statistically comparable. Notably, approach 2 exhibited occasional outliers with more intensive steering corrections, suggesting that these drivers may occasionally require stronger adjustments.

For left steering events, the analysis was limited by small amount of events in approach 2 ($n = 7$) and approach 3 ($n = 10$), reducing the precision and representativeness of the findings. Approach 2 demonstrated a higher average event rate, with drivers averaging 160 km per event, compared to 120 km per event for approach 3. This indicates more frequent left steering events in approach 2, potentially reflecting differences in driving conditions or training focus. The mean left steering acceleration differed, with approach 2 recording a mean of -3.25 m/s$^2$ and a $SD = 0.34$, while approach 3 exhibited a mean of -3.42 m/s$^2$ with the same SD. These results suggest that while approach 3 drivers displayed slightly more consistent steering behavior, the variability in approach 2 was higher, potentially reflecting more dynamic cornering such as having a more aggressive driving style.

Statistical tests support the comparability of left steering events between the groups. The Mann-Whitney \mbox{U-test} resulted in a p-value of $p= 0.764$, and Levene’s test reported $p = 0.183$, indicating no significant differences in distribution or variability despite the differences in event frequency.

While approach 3 shows a higher frequency of right-hand steering events and approach 2 demonstrates more frequent left-hand steering events, both groups exhibit comparable distributions and variability in steering accelerations. The small sample sizes, particularly for left steering events, limit the statistical power of these results, reducing their precision and representativeness. Nonetheless, the findings suggest that the observed differences in steering frequency are more likely influenced by environmental and operational conditions than by fundamental differences in steering behavior between the two training approaches.

\subsubsection{Summary of detailed vs. scenario-based ODD Training Results}
The experimental results show that approach 2 demonstrates a more controlled and anticipatory driving style, particularly in braking, where significant differences were observed. Results for other event types do not show statistically significant differences, suggesting that the disparities between the groups are primarily rooted in braking strategy and control. 

The intensive training of approach 2, which encompasses a wide range of possible conditions and parameters within a specific ODD, enables drivers to develop a deeper understanding of various factors influencing driving behavior. This type of training allows them to drive more predicatively and with greater control in diverse scenarios, which could be reflected in longer durations and distances until the next braking event. The ability to anticipate potential scenarios and respond accordingly may result in approach 2 needing to brake or steer abruptly less frequently, as they are more familiar with and adept at assessing influencing variables. Considering these aspects, approach 2 does not appear to present safety-critical concerns.

The training of approach 3 focuses on representative scenarios that reflect typical challenges within an ODD but does not address all potential variables in detail. This means that while drivers are well-prepared for common and realistic situations, they may lack specific knowledge about rare or extreme conditions. In typical driving situations, their performance is solid, as indicated by the longer intervals observed for acceleration and right-steering events. However, their ability to respond to more complex or unforeseen scenarios might not match the depth achieved by approach 2, which could potentially lead to a higher frequency of braking maneuvers. Further, the observed significance's could possibly be explained by the distribution of the RD driving experience as well. While the results do not provide direct evidence of safety-critical behavior, the reduced preparation for worst-case scenarios might increase the safety risk if unusual situations arise.

\subsection{No Training vs. specific ODD Training}
The necessity of ODD-specific training for remote driving in public road contexts was evaluated by comparing RDs, which were not trained for a defined ODD, with those specifically trained for a defined ODD. The goal was to assess whether RDs with no ODD-specific training exhibit significant behavioral abnormalities or safety-critical deficits compared to their trained counterparts, thus providing insight into the importance of such training for RD performance and safety.

\begin{figure} [h!]
    \centerline{\includegraphics[width=3.4in]{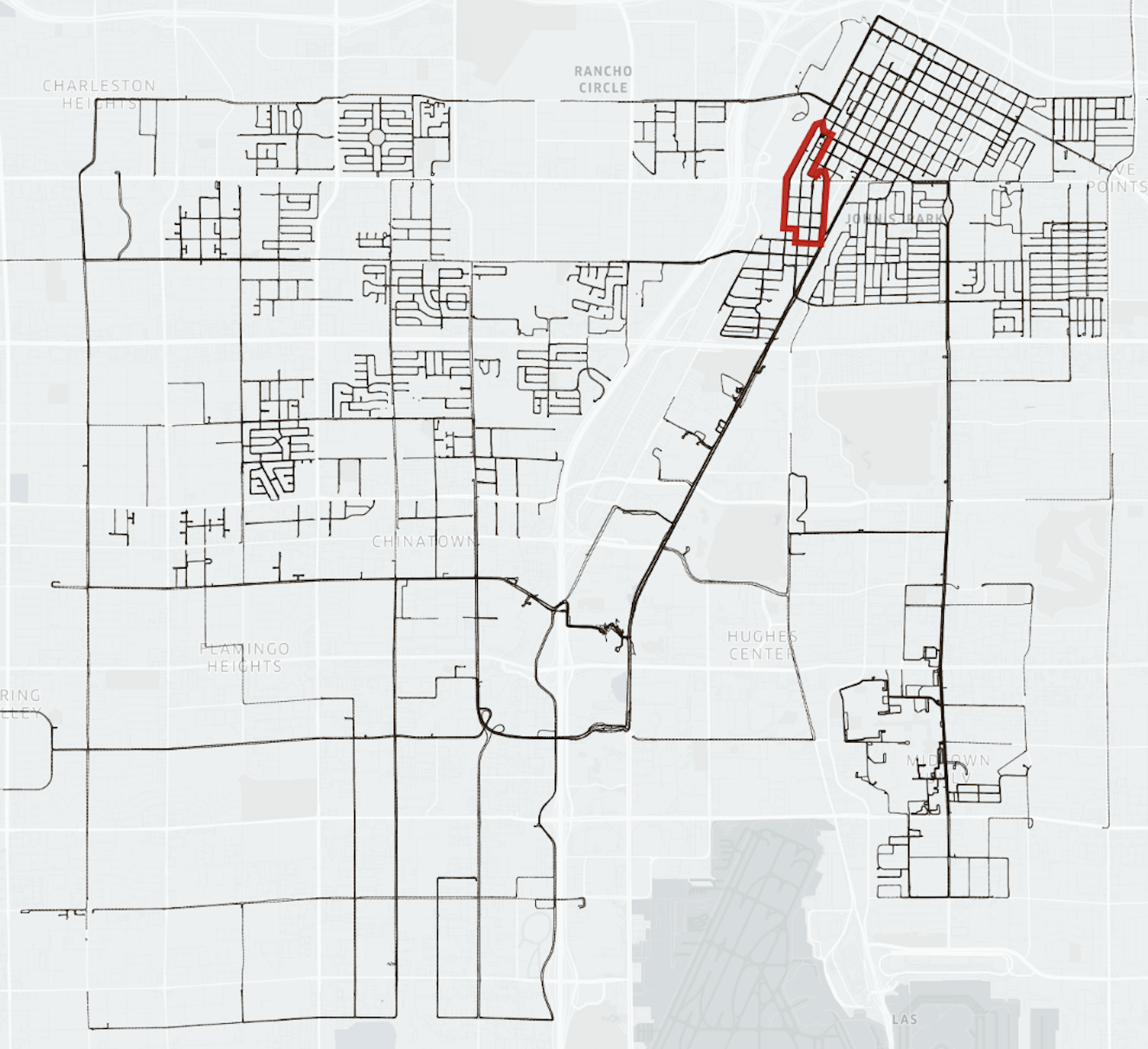}}
    \caption{Geographical part of the Operational Design Domain (ODD) in Las Vegas, Nevada, US. Red = Geographical ODD part of approach 2 and 3 as shown in Fig. \ref{fig2}; Black = Remote driving sessions of approach 1.}
    \label{Keplermapgroup1}
\end{figure}

\begin{itemize} 
    \item \textbf{Untrained RDs (Approach 1):} RDs in this group were not specifically trained in an ODD-based approach. They were exposed to general driving scenarios and conditions as defined in Section IV.\ref{RDTraining} without any targeted preparation for specific operational domains. This group consists of 10 RD, aged between 29 and 33 years (\mbox{$M = 30.67$}, \mbox{$SD = 1.86$}), where 7 identified themselves as male and 3 as female. The group covered a total of about 1670 km of streets, with the ODD, visualized in Fig. \ref{Keplermapgroup1}, covering a street length of 1767 km, from \mbox{August 1}, 2023, to April 30, 2024.
    \item \textbf{ODD-specific trained RDs (Combination of approach 2 and 3):} RDs in this group received specific training for ODDs (scenario-based or detailed road section-based ODD training), specifically tailored for public road contexts. In total 13 RD are within this training group, aged between 24 and 33 years (\mbox{$M = 29.44$}, \mbox{$SD = 2.79$}), 9 identified themselves as male and 4 as female. Over the same time period from \mbox{August 1}, 2023, to April 30, 2024, this group drove about \mbox{1833 km} of streets within the ODD shown in Fig. \ref{fig2} covering a street length of 9 km, focusing on enhancing safety and performance in specific road types. 
\end{itemize}

By comparing these approaches, the analysis aims to determine if ODD-specific training provides a significant safety advantage and reduces safety-critical deficits in remote driving performance. Compared to Section VI.\ref{detailedvsscenario}, the distribution of RD experience between the two approaches is comparable as shown in Tab. \ref{RDbalance}, so that the effects regarding the driving experience in kilometers from Section \ref{Investigation} have no effect on this comparison.

\begin{table}[ht]
\centering
\begin{tabular}{p{45pt}p{65pt}p{75pt}}
\hline
RD experience & No ODD-specific training & ODD-specific training \\ \hline \hline
$<$ 500 km & 823.37 km & 940.20 km \\ \hline 
$>$ 500 km & 846.49 km & 892.77 km \\ \hline
\end{tabular}
\caption{Overview of Remote Driver (RD) experience for approach 1 (No ODD-specific training) and combined approach 2/3 (ODD-specific training).}
\label{RDbalance}
\end{table}

\subsubsection{Longitudinal Event Results}
\begin{figure}[!h]
\centerline{\includegraphics[width=3.5in]{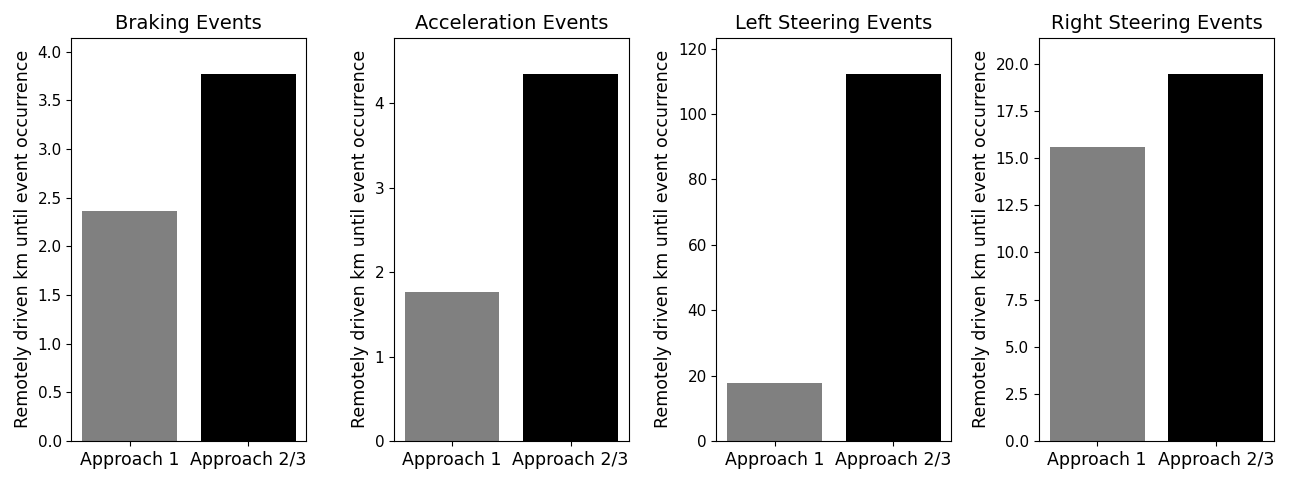}}
\caption{Remotely driven kilometers per event group until event occurrence for approach 1 and the combined approach 2 and 3.\label{kmperevent123}}
\end{figure}

The analysis of braking events highlights a distinct difference in driving style between the two approaches. Drivers in approach 2/3, who underwent ODD-specific training, demonstrated a more anticipatory driving style characterized by less frequent braking maneuvers as visualized in \mbox{Fig. \ref{kmperevent123}}. However, the overall braking performance, as reflected by median deceleration values, remained similar between the approaches. Approach 1 had a mean deceleration of \mbox{-2.74 m/s$^2$} ($SD = 0.69$), while approach 2/3 recorded a slightly lower mean of \mbox{-2.70 m/s$^2$} ($SD = 0.85$).

Statistical tests confirm that these differences in braking metrics are not statistically significant. The Mann-Whitney \mbox{U-test} yielded a p-value of $p = 0.153$, while Levene's test for equality of variances reported a p-value of $p = 0.152$, indicating comparable distributions in braking performance across the approaches. These results suggest that while the ODD training influenced the frequency of braking events, the magnitude of braking force remained consistent.

For acceleration events, approach 2/3 also exhibited clear distinctions in driving behavior compared to approach 1. In approach 2/3 drivers harshly braked less frequently and demonstrated more stable and consistent acceleration patterns due to the higher amount of kilometers until an braking event occurs. The mean acceleration for approach 2/3 was 2.09 m/s$^2$ ($SD = 0.14$), reflecting a controlled driving style. Conversely, approach 1 recorded a higher mean acceleration of \mbox{2.14 m/s$^2$} with a greater variability ($SD = 0.26$).

\begin{figure}[!h]
\centerline{\includegraphics[width=3.5in]{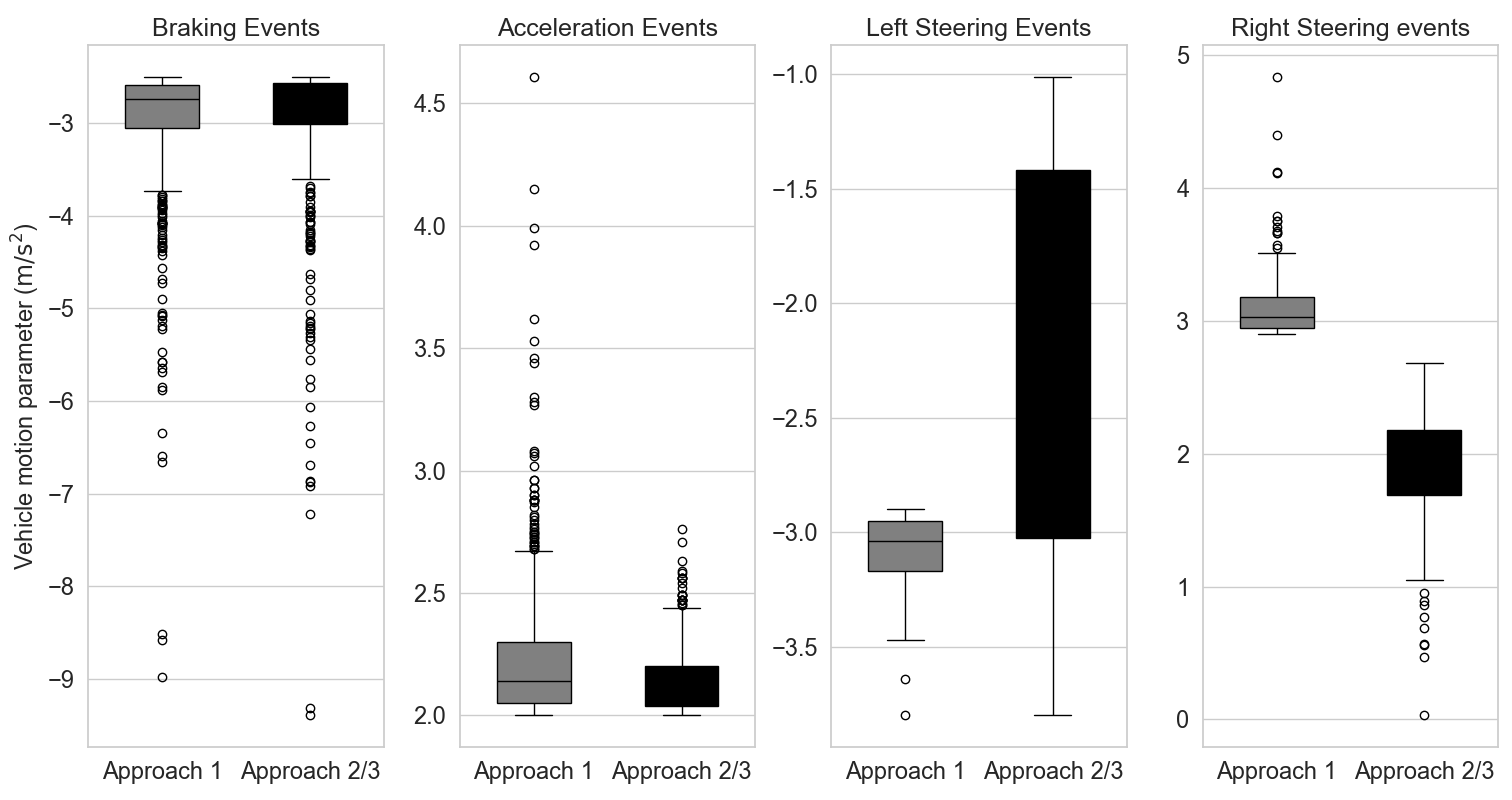}}
\caption{Driving performance boxplots per event group for approach 1 and the combined approach 2 and 3.\label{boxplotperevent123}}
\vspace{-2mm}
\end{figure}

Statistical analysis revealed significant differences between the groups in the distribution and variance of acceleration events ($p < 0.01$). Approach 1 exhibited more extreme outliers in acceleration values as shown in \mbox{Fig. \ref{boxplotperevent123}}, indicating a less controlled driving style. These findings suggest that ODD-specific training in approach 2/3 not only reduces braking frequency but also fosters a smoother and more predictable driving performance.

\subsubsection{Lateral Event Results}
The analysis of left steering events reveals that \mbox{approach 2/3}, with ODD-specific training, performed significantly fewer left steering maneuvers compared to approach 1 (see \mbox{Fig. \ref{kmperevent123}}). On average, approach 2/3 covered 110 km per event, while approach 1 covered only 19 km per event, indicating a less frequent need for corrective or extreme steering maneuvers in approach 2/3. The median steering values further underscore this difference. Approach 1 had a median of -3.04 m/s$^2$, reflecting more extreme steering to the left, whereas approach 2/3 exhibited a less pronounced median of -1.63 m/s$^2$. This suggests that approach 2/3 employed smoother and less abrupt steering corrections.

Statistical tests confirmed significant differences in the distribution and variance of left steering events ($p < 0.01$). These results highlight the impact of ODD-specific training in fostering more controlled lateral driving behaviors.

For right steering events, the differences between the approaches were less pronounced compared to left steering. Both approaches exhibited similar distances covered per event, with approach 1 showing a slightly higher frequency of right steering maneuvers (see \mbox{Fig. \ref{kmperevent123}}). Approach 1 also displayed a greater tendency for variability in right steering, with some values exceeding 4 m/s$^2$, marked as outliers. These extreme values were not present in \mbox{approach 2/3} as shown in Fig. \ref{boxplotperevent123}, indicating a more consistent and stable lateral driving pattern among the ODD-trained drivers.

As with left steering, significant differences in the distribution and variance of right steering events were confirmed ($p < 0.01$). Overall, the results suggest that while the impact of ODD-specific training is less pronounced for right steering events, approach 2/3 still benefits from a more uniform and predictable driving style compared to approach 1.

\section{LIMITATIONS}
\label{Limitation}
This study has several limitations that should be acknowledged and discussed below.

Firstly, the effects observed in this study are based on a small RD sample size of $n = 14$. As a result, outliers have a stronger influence on the results, which increases the width of the confidence intervals. This limitation may affect the precision and generalizability of the findings.

Secondly, the results are specific to the system and ODD defined for this study with a particular use case. While they provide valuable insights, they may not fully translate to other vehicle platforms such as trucks or other ODDs such as desert environments without further validation. However, these findings can serve as an indication for similar applications in different ODDs or vehicle platforms, providing a foundation for further research and adaptation.

Thirdly, the analysis does not consider individual RD parameters such as situational awareness, stress resilience, and workload levels. These factors are known to significantly influence driver decision-making and behavior during remote driving tasks. By neglecting these individual parameters, the study may miss important insights into the specific challenges and complexities faced by drivers in real-world scenarios. This limitation restricts a deeper understanding of the human factors that affect remote driving performance.

Finally, the comparison between the ODDs used in approach 1 and the combined approach 2/3 takes place at the abstraction level of logical driving scenarios \cite{wachenfeld2016safety}. The ODD in approach 1 is geographically larger, which increases the likelihood of event clustering in shorter road sections, especially where long, uninterrupted streets are less common. Although the scenarios themselves are similar on a logical level, differences in traffic density and road design may influence event frequency, potentially skewing the results. Given the limited number of RDs, some residual variability due to scenario differences cannot be fully ruled out. This distinction highlights the challenges in directly comparing approaches across ODDs with varying geographical and environmental characteristics. However, the large distance driven in this study ensures broad coverage of a wide range of scenarios and conditions, which statistically levels out short-term or situational influences such as specific environmental or traffic conditions. 

\section{CONCLUSION}
\label{conclusion}
The findings demonstrate that increased driving experience significantly enhances vehicle control, with noticeable improvements in reducing abrupt steering, braking, and acceleration behaviors up to approximately 600 km of cumulative driving. The slight increase in these events between 800 and 1000 km suggests a plateau in the learning curve, reflecting routine development and the solidification of driving habits. Correlation analyses further substantiate this trend, highlighting the progressive learning curve as drivers adapt to the remote driving system and their ODD, which is also confirmed by the analysis from Hans and Adamy \cite{hans2025identification} focusing on the identification and classification of human performance challenges from real-world remote driving data in urban ODDs. 

In addition to that, remote driving efficiency exhibited a positive trend with increasing kilometers, likely indicating adaptation to the system and greater consistency in driving behavior, particularly during the first 300 km of experience. From approximately \mbox{400 km} onward, RD efficiency plateaued within a range of 0.35 to \mbox{0.42 km/min}, suggesting that cumulative experience led to consistent and effective vehicle operation. However, variability among drivers, as observed in the confidence intervals, underscores individual differences in driving styles, abilities, and the degree to which cumulative experience influences performance. This variability was further supported by the correlation between efficiency and km/day, indicating that with increased driving experience, efficiency became less dependent on daily kilometers driven or time spent driving.

In the second part of this study, the comparison of particular ODD training approaches revealed notable advantages for ODD-specific training. Drivers who received targeted ODD-specific training, demonstrated greater control and stability, with longer intervals between critical events and reduced variability in acceleration and steering. These findings suggest a more anticipatory and less interventionist driving style, critical for safety in RDSs. In contrast, the training approach, which lacked ODD-specific training, exhibited higher variability and a greater frequency of safety-critical interventions, confirming the effectiveness of tailored ODD training in preparing drivers for complex scenarios.

Further distinctions were observed between the detailed road-section-based ODD training and scenario-based ODD training. While detailed ODD training comprehensive preparation resulted in high robustness and resilience, scenario-based approach proved sufficient for typical challenges but may carry slightly higher risks in extreme or unexpected situations. A hybrid solution combining scenario-based training with detailed preparation for specific, high-risk areas such as the Las Vegas Strip emerged as the most promising approach. This balance ensures scalability while addressing critical edge-case scenarios, supported by subjective driver feedback emphasizing the practicality of adaptable scenario-based training alongside detailed training for specialized conditions. These insights underline the importance of driving experience for RD and ODD-specific training for improving safety, efficiency, and stability in RDSs, paving the way for optimized training protocols that balance scalability with thorough preparation for high-risk scenarios.

For future work a detailed analysis of the specific scenarios is essential, with particular attention given to the human factor. This will help evaluate the issues and identify the underlying reasons that led to the observed events. Such an in-depth investigation could provide valuable insights into the cognitive processes of the RD, as well as potential technical or operational limitations that may have influenced system performance. In addition, further research should evaluate if a hybrid approach of the scenario-based ODD training approach and the detailed training approach in special areas could improve situational awareness, decision-making, and adaptability, ultimately leading to more effective and safer remote driving performance. Both would be crucial in developing strategies to optimize both the performance and safety of RDS.

\section{ACKNOWLEDGMENTS}
This study was supported by Vay Technology, a company operating a remotely driven commercial car-sharing service without a safety driver in Las Vegas, Nevada, US. The authors would like to express sincere gratitude to Athanassios Lagospiris, Mathias Metzler, and Hans-Leo Ross for their invaluable support, insightful input, and constructive comments on this paper. Their contributions were instrumental in shaping the direction and findings of this research.

%% Define the bibliography file to be used
\bibliography{Hans_biblio}

\newpage

\begin{IEEEbiography}[{\includegraphics[width=1in,height=1.25in,clip,keepaspectratio]{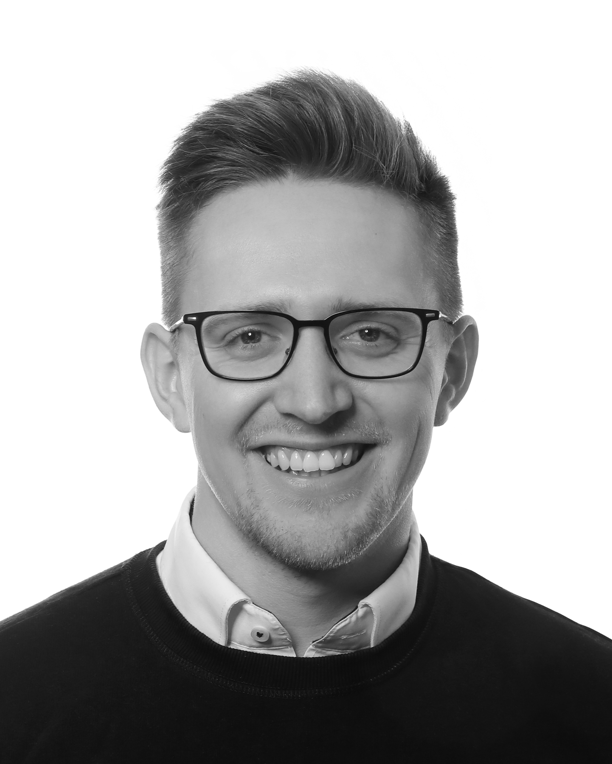}}]{OLE HANS } received the B.Sc. degree in Safety Engineering from University at Wuppertal in 2020 and the M.Sc. degree from University at Wuppertal in Quality Engineering in 2022. He is currently pursuing the Ph.D. (Dr.-Ing.) degree in electrical engineering and information technology with the Institute of Automatic Control and Mechatronics at the Technical University of Darmstadt. His research interests include safety of remotely driven and automated vehicles.
Since 2022, he has also been part of the Operational Safety Department at Vay Technology and is responsible for the safety-related operation of remote-driven vehicles.\end{IEEEbiography}

\begin{IEEEbiography}[{\includegraphics[width=1in,height=1.25in,clip,keepaspectratio]{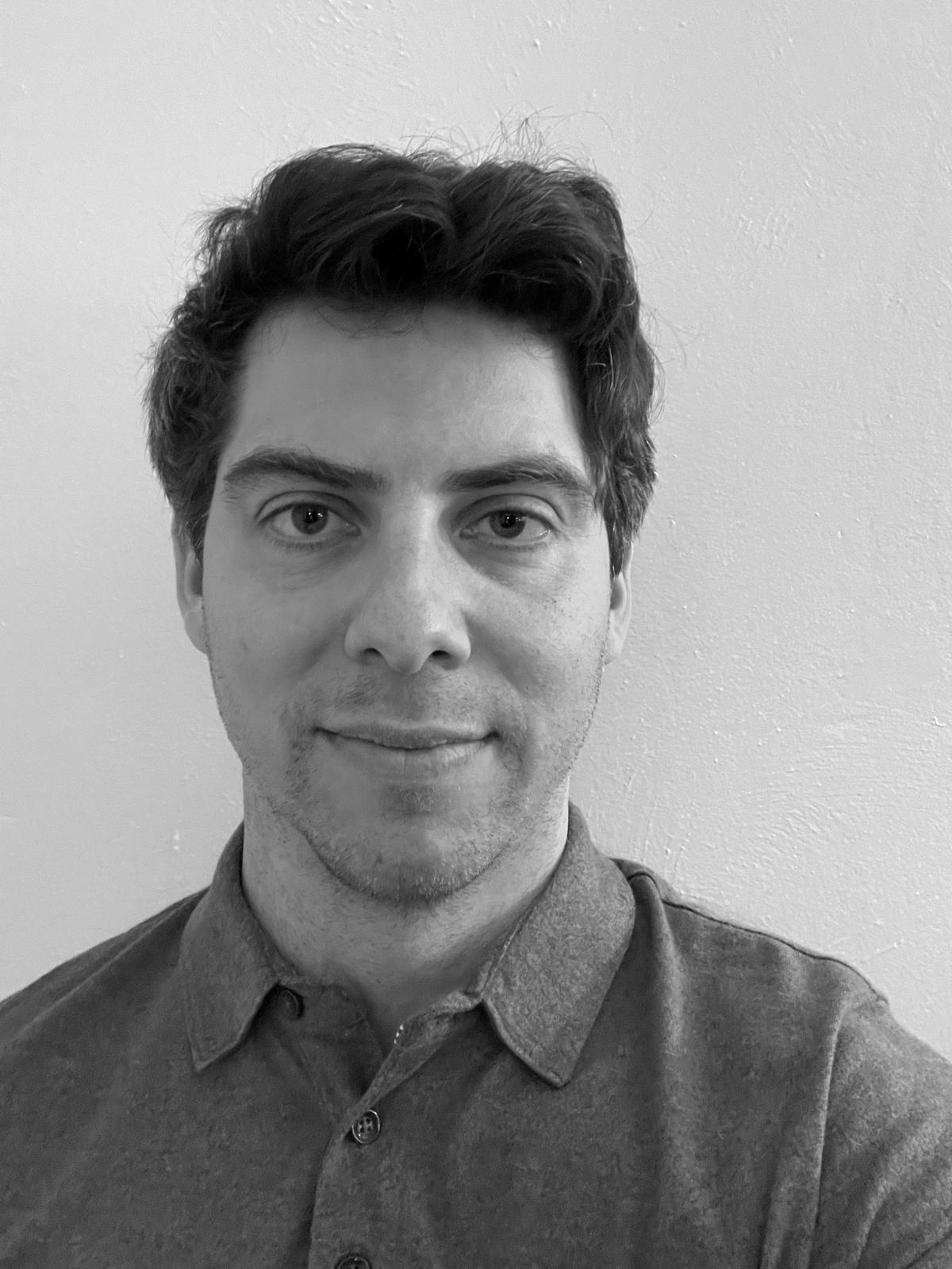}}]{BENEDIKT WALTER } received his B.Sc. and M.Sc. in aerospace engineering from the University of Stuttgart, Germany in 2013, respectively 2015. Further he received a Ph.D. (Dr.-Ing.) from the University of Stuttgart in systems engineering and computer science in 2020. His research focus is on system design and knowledge representations for automotive systems.
From 2015 to 2018 he worked as a Ph.D. student at Mercedes-Benz in a collaboration with the University of Stuttgart. During the time from 2018 to 2022 he worked as the Program Manager of the ADAS and ADS software development project at Mercedes-Benz. 2022 and 2023 he worked at US Safety Council at NIO. Between 2023 and 2024 he was part of the Operational Safety Department at Vay Technology where he contributed to this work. Since 2025 he is Staff Systems Engineer at Torc Robotics.\end{IEEEbiography}
\begin{IEEEbiography}[{\includegraphics[width=1in,height=1.25in,clip,keepaspectratio]{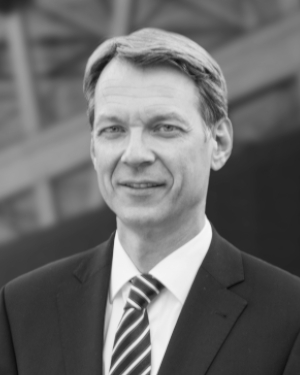}}]{JÜRGEN ADAMY }
received the Diploma and Dr.-Ing. degrees in Electrical Engineering from the Technical University of Dortmund, Germany, in 1987 and 1991, respectively. From 1992 to 1998, he was an Engineer and a Manager in the area of Control Applications at Siemens AG, Erlangen, Germany. In 1998, he became a Full Professor at the Technical University of Darmstadt, Germany, where he is the head of the Control Methods and Intelligent Systems Laboratory. His research interests are in the areas of nonlinear control, intelligent systems, and mobile robots.\end{IEEEbiography}

\end{document}